\documentclass[manuscript]{aastex}
\usepackage{graphicx}
\usepackage{epstopdf}

\shorttitle{Models for 15 Galactic Supernova Remnants}
\shortauthors{Leahy \& Ranasinghe}

\begin{document}


\title{Evolutionary Models for 15 Galactic Supernova Remnants with New Distances}


\author{D.A. Leahy and S. Ranasinghe}
\affil{Department of Physics $\&$ Astronomy, University of Calgary, Calgary,
Alberta T2N 1N4, Canada}



\begin{abstract}
{Recent studies using 21 cm HI line and $^{13}$CO line observations in the inner part of the Galaxy have resulted in new distances for 30 Galactic supernova remnants (SNRs).
15 of those remnants have observed X-ray spectra, for which shocked-gas temperatures and
emission measures are measured. 
Here we apply spherically symmetric SNR evolution models to these 15 remnants 
to obtain estimates for  ages, explosion energies, circum-stellar medium densities and profiles
(uniform or wind-type).
From the distribution of ages we obtain a supernova birth rate and estimate incompleteness. 
The energies and densities can be well fit with log-normal distributions. 
The distribution of explosion energies is very similar to that of SNRs in the Large
Magellanic Cloud (LMC), suggesting SN explosions in the LMC and in the Galaxy are very similar.
The density distribution has higher mean density for Galactic SNRs than for LMC SNRs, by a factor $\sim$2.5.}

\end{abstract}





\section{Introduction}

The study of supernova remnants (SNRs) is of great interest in astrophysics
(see \cite{2012Vink} and references therein for a recent review). 
SNRs provide valuable information relevant to stellar evolution, the evolution of the Galaxy and its interstellar medium. 
SNRs are the dominant source of kinetic energy input into the interstellar medium \citep{2005Cox} and thus measuring SNR energetics is critical to understanding the structure of the interstellar medium. 

SNRs are observed primarily in X-rays, by emission from hot interior gas with 
temperature $\sim$1 keV, and in radio, by synchrotron emission from 
relativistic electrons accelerated by the SNR shockwave. 
The observational constraints for different SNRs are often different in nature. 
They depend on the brightness of emissions in different wavebands by a given 
SNR and by the instruments used to observe that SNR.
Only a small fraction of the $\sim$300 observed SNRs in our Galaxy have previously
been well enough characterized to determine their evolutionary state, including explosion 
type, explosion energy and age. 
A few historical SNR have been observed in great detail and modelled with hydrodynamic simulations. 
For example, Tycho has been modelled \citep{2006Badenes} and used to test different
models for SN Type Ia explosions.
However, most SNRs are not observed nearly as well and have not been subject to similar
detailed modelling.
For these observationally less-constrained SNRs, it is worthwhile to determine their bulk physical characteristics,
but with a simpler approach than full hydrodynamic modelling.

In order to expedite characterization of SNRs, we have developed a set of 
analytical SNR models for spherically symmetric SNRs and implemented them in Python \citep{2017LeahyWilliams}. 
The set of models includes a wide set of models constructed previously by other authors. 
We carried out the additional step of consistently joining different stages of evolution, which in several cases has not been done before. 
The resulting models facilitate the process of using different constraints 
from observations to estimate SNR physical properties of interest.

The current work includes the following. 
First, we extend the models to calculate the X-ray emission from a SNR during 
the evolutionary phase between the self-similar ejecta dominated phase and the self-similar Sedov-Taylor phase.
Then we solve the inverse problem of how to determine the initial
parameters of a SNR using its observed properties.
Finally, we apply the solution of the inverse problem to a set of SNRs with newly determined distances.
The structure of the paper is as follows.
Section~\ref{sec:overview} presents an overview of the SNR models and the solution of the inverse problem. 
%The new models include the important ejecta-dominated to adiabatic phase transition.
%We discuss how to solve the inverse problem of finding initial conditions from
%the current state of the SNR. 
Section~\ref{sec:sample} describes the supernova remnant sample and the model fits for individual SNRs.
Section~\ref{sec:properties} present our analysis of the properties of the SNR sample and Section~\ref{sec:conclusion} summarizes the results on the distribution of explosion energies and densities.

\section{Overview of the Supernova Remnant Evolution}\label{sec:overview}

\subsection{General SNR Evolution}

A SNR is, by definition, the interaction of the SN progenitor's ejecta with the interstellar medium (ISM) or circumstellar  medium (CSM). 
The various stages of evolution of a SNR include, in order, the 
ejecta-dominated stage (labelled ED), the adiabatic (or Sedov-Taylor) stage
(labelled ST), and radiative stages.
These are reviewed in part in many previous publications, including \citet{2017LeahyWilliams}, \cite{1988cioffi}, \cite{1999truelove} (hereafter TM99) and \citet{2012Vink}.
The radiative stages are usually divided into the earlier pressure driven
snowplow (PDS) stage and later momentum conserving shell (MCS) stage.
In addition to these phases, there are the transitions between stages,
which we label as ED to ST, ST to PDS, and PDS to MCS, respectively.
The ED to ST stage is particularly important, because the SNR is still bright in
X-rays and radio, and it is long-lived enough that a significant number of Galactic
SNRs could be in this phase.
The end of the life of the SNR is usually taken to be when the SNR merges
with the ISM, i.e. the shock velocity has dropped to a value similar to the
random velocities in the general ISM.
 
\subsection{The Adopted Model for SNRs}

For simplicity we assume that the SN ejecta and CSM or ISM are spherically symmetric.
The required hydrodynamic equations for a spherically symmetric SNR evolution are summarized by \citet{1991WL}.
The ISM/CSM density profile is taken as a power-law centred on the SN explosion $\rho_{CSM}=\rho_s r^{-s}$, with s=0 (constant
density medium) or s=2 (stellar wind density profile).
The unshocked ejectum is taken to have a power-law density profile $\rho_{ej}\propto r^{-n}$.
For these assumptions, the early part of the SNR evolution (the ED phase) 
follows a self-similar evolution, as determined by \cite{1982Chev} and \cite{1985Nad}.
The SNR shock radius evolution was extended for the ED through ST phases by TM99.

%\subsubsection{General Aspects of the Model}
The model for SNR evolution that we construct is partly based on the TM99 analytic solutions, with  additional
features. % described here.
A detailed description of the most of the model is given in \citet{2017LeahyWilliams}. 
For example, we used Coulomb collisional electron heating, which is consistent with the results of \cite{2013Ghav}. %Ghavamian et al. 2013.
However, important extensions have been added in the current work. 
The main extensions are:  i) we calculate the emission measure and
emission measure-weighted temperature during the transition from ED phase to ST phase;  ii) the inverse problem is solved, 
which takes as input the SNR observed properties and determines the 
initial properties of the SNR.

The SNRs modelled here have relatively small physical diameters, thus they are generally in
a stage prior to the radiative stage. 
We checked after obtaining our model results that none of the models had reached the radiative phase,
as defined by the time $t_{pds}$ (for example, see TM99).
Such non-radiative supernova remnants follow unified evolution (TM99). 
For the case of uniform ISM, the characteristic radius 
is $R_{ch}=(M_{ej}/\rho_0)^{1/3}$ 
and characteristic time is $t_{ch}=E_0^{-1/2}M_{ej}^{5/6}\rho_0^{-1/3}$. 
Here $M_{ej}$ is the mass of ejecta, $E_0$ is the explosion energy, 
$\rho_0$ is the mass density of the ISM, $\rho_0=\mu_{H} n_{0} m_{H}$, 
and  $n_0$ is the hydrogen number density of the ISM.
The characteristic velocity is $v_{ch}=R_{ch}/t_{ch}$ and the characteristic
shock temperature  is $T_{ch}=\frac{3}{16}\mu_I\frac{m_H}{k_B}v_{ch}^2$.
For the case of a circumstellar wind medium, $\rho \propto r^{-s}$, s=2, 
the  characteristic radius and time are 
$R_{ch}=(M_{ej}/\rho_s)$ and $t_{ch}=E_0^{-1/2}M_{ej}^{3/2}v_w/\dot{M}$.
Here $\dot{M}$ and $v_w$ are the wind mass loss rate and velocity, 
yielding $\rho_s=\frac{\dot{M}}{4\pi v_w}$.

For SNRs which are less than several thousand years old, 
the post shock gas is a mixture of electrons of lower temperature and different species of ions, each with their own (higher) temperature.
The mechanism of electrons heating is taken to be Coulomb collisions.
As shown by  \citet{2017LeahyWilliams}, this electron heating model agrees well with the observed contraints on electron heating \citep{2013Ghav}.
This allows calculation of the electron temperature as a function of time,
so that the X-ray (electron) temperature can be determined in the model.
%For details see \citet{2017LeahyWilliams}.
%definition of the shock temperature, which is the temperature to which a 
%(fictitious) species of mass $\mu_{tot}m_H$ would be heated, with
%$1/\mu_{tot}=1/\mu_{e}+1/\mu_{ion}$.

Element abundances for the ISM/CSM are taken to be solar values, from \cite{1998GS}.
%Solar: 12,10.93,8.52,7.92,8.83,8.08,7.58,7.55,7.33,7.50) by number
%LMC: 12,10.94,8.04,7.14,8.35,7.61,7.47,7.81,6.70,7.23)
For the SN ejecta, generic abundances values were taken (in units of $log(X/H)+12$) 
for massive star core-collapse SN as:
 He: 11.22, C: 9.25, N: 8.62, O: 9.69, Ne: 8.92, Mg: 8.30, Si: 8.79,
 S: 8.54, Fe: 8.55; 
and for Type Ia SN as:
 He: 11.40, C: 12.60, N: 7.50, O: 12.91, Ne: 11.04, Mg: 11.55, Si: 12.75,
 S: 12.43, Fe: 13.12.
The hydrodynamic boundary conditions at the contact discontinuity involve pressure, 
mass density, velocity and temperature, whereas the EM depends on electron
and ion number densities. 
These are related by $\rho=\mu_e m_{H} n_e=\mu_{ion} m_{H} n_{ion}$, thus the composition 
affects the temperature and emission measure of the reverse shock gas.
14 of the 15 SNR in our sample have been previously classified as core-collapse or unknown type. 
 Thus for simplicity,  we use the ejecta abundances for core-collapse SN. 

The evolution starts with the early ED self-similar phase (valid for $t<t_{core}$ in the notation of TM99).
For ejecta with $n>5$ we use the self-similar solutions given by \cite{1982Chev}. 
For the current modelling we restrict our calculations to $n>5$
and either $s=0$ (uniform ISM) or $s=2$ (stellar wind CSM).

During the adiabatic (ST) phase valid for $t>t_{rev}$ (using the notation of TM99), 
the shock radius and shock velocity evolution were calculated using the TM99 solutions. These are much more accurate
than the simple Sedov equations, as shown by \citet{2017LeahyWilliams}. 
The reason is that the Sedov models do not include the effect of ejected mass on changing the timescale of evolution.

The transition period from ED phase to ST phase occupies a significant amount of time (TM99). 
For example, for n=7, the ED phase ends at $t_{cor}=0.363 t_{ch}$, the transition
phase lasts from $t_{cor}$ to $t_{rev}=2.69 t_{ch}$: the transition
phase is 7.4 times as long as the self-similar phase.
The shock radius and shock velocity during the transition were calculated using the results of TM99. 

For the $s=2$ case, the period of self-similar evolution is much longer (e.g., TM99).
Thus we only use the self-similar ED phase for $s=2$. 

\subsubsection{Emission Measures and Temperatures for Ejecta-Dominated, Transition and Adiabatic Phases}

The radiation from the SNR post-shock plasma occurs via two body processes, e.g. \citet{1976Ray}.
Thus it depends on the emission measure, $EM$, $EM=\int n_e(r) n_H(r) dV$
with electron density $n_e$ and hydrogen density $n_H$.
$EM$ is measured for a given SNR from its X-ray spectrum, with: 
$norm=\frac{10^{-14}}{4\pi D^2}EM$, with $D$, the distance to the SNR 
with units in cgs (XSPEC manual at heasarc.gsfc.nasa.gov).
The $EM$ is calculated for a given model using the interior density profile of that SNR model. 
We define the dimensionless $EM$, 
%\footnote{The dimensionless emission measure, dEM, is not to be confused with differential emission measure, which is often written as DEM.}
$dEM$, by
\begin{equation}
dEM=EM/(n_{e,s}n_{H,s}R_s^3)
\end{equation}
with $n_{e,s}$ and $n_{H,s}$ the values  %of $n_e$ and  $n_H$ 
just inside the shock front. 
$dEM$ depends only on the profile of the interior density distribution.

The observed temperature of a SNR from the X-ray spectrum 
is the emission-weighted temperature of the interior gas:
\begin{equation}
T_{EM}=\frac{1}{EM}\int  n_e(r) n_H(r) T(r) dV
\end{equation}
We define the dimensionless emission-weighted temperature 
as the emission-weighted temperature divided by the shock temperature, $T_s$: $dT=T_{EM}/T_s$.
%so that the emission-weighted temperature is $dT\times T_s$.
$dT$ depends on the interior profiles of density and of temperature. 

We distinguish between forward-shocked (hereafter fs) and reverse-shocked (hereafter rs) interior SNR gas
and define the corresponding dimensionless $EM$'s ($dEM_{fs}$ and $dEM_{rs}$)
and the dimensionless emission-weighted temperatures  ($dT_{fs}$ and $dT_{rs}$).
From the observed X-ray spectra it is sometimes clear, based on the observed element abundances,
whether the emission is from fs gas or from rs gas.
In such cases, the appropriate model emission (fs or rs) is used. 

The interior structure ($T(r)$ and $\rho(r)$ at any given time) is required 
to calculate $dEM_{fs}$, $dEM_{rs}$, $dT_{fs}$ and $dT_{rs}$.
$dEM$ and $dT$ are independent of time for the ED and ST self-similar phases of evolution, for which the density profiles have
a constant shape. 
The interior structure solutions for the ED phase for different values of s and n are given by the solutions in \citet{1982Chev}.
We use those solutions to calculate $dEM_{fs}$, $dEM_{rs}$, $dT_{fs}$ and $dT_{rs}$.

The SNR interior structure for the fs gas in the ST phase was calculated using solutions to the hydrodynamic 
equations carried out by \citet{2017LeahyWilliams}.
During the ST phase, the fully shocked ejecta has a density profile peaked at the 
contact discontinuity and is expanding slowly. 
\citet{1988cioffi} shows interior density profiles for a SNR with ejecta during the ST phase and later.
The ejecta expansion speed at the contact discontinuity (at position $r_{cd}$) is  $\sim V_{fs} r_{cd}/R_{fs}$.
%where $ V_s$ and $R_s$ are the forward shock velocity and radius.
We carried out numerical solutions to the hydrodynamic equations with ejected mass for the s=0, n=0 case to obtain 
 $dEM_{rs}$ and $dT_{rs}$ in the ST phase.

To calculate $dEM_{fs}$, $dEM_{rs}$, $dT_{fs}$ and $dT_{rs}$ for the ED to ST transition, 
we used  hydrodynamic simulations to obtain the SNR interior profiles.
These showed that $dEM$ and $dT$ vary smoothly with time between the ED and ST phases for both
fs ISM and rs ejecta.
Here we use an interpolation, as a function of  $t/t_{ch}$, to obtain the values
 $dEM$ and $dT$ for transition period between the ED phase and the ST phase.
From the hydrodynamic simulations, we find that the interpolation is a good approximation ($\sim5\%$ error). 
This is similar to the error in using the TM99 analytic approximations for the shock evolution.

\subsubsection{The inverse problem: application of models to SNR data}

The forward modelling, taking initial SNR conditions and calculating conditions at time t, is described in \citet{2017LeahyWilliams}.
The input parameters of the SNR forwardnmodel are time (in yr) plus the following.
Ejecta parameters are:  energy $E_0$; ejected mass $M_{ej}$; ejecta power-law index n; and ejecta composition.
The ISM/CSM parameters are: temperature $T_{ISM}$; power-law index s (0 or 2); 
density $n_0$ (if s=0) or mass-loss parameter $\rho_s=\frac{\dot{M}}{4\pi v_w}$ (if s=2);
and ISM/CSM composition.
The total number of parameters is 7, taking the composition of ISM/CSM and ejecta to be fixed.  
For the Galactic SNRs modelled here, we take the ISM/CSM to be solar composition. 

For the inverse model, the input  parameters are normally the current fs radius,
$EM$ and $T_{EM}$. The outer radius of the SNR is the fs radius.
$EM$ and $T_{EM}$ are usually dominated by fs emission.
However in cases where the element abundances are measured to be enhanced, the emission is %$EM$ and $T_{EM}$
dominated by rs ejecta or from a combination of fs ISM and rs ejecta.  
%In applying the model, we test these different cases because it is often not known in advance which region dominates the X-ray emission.
In some cases both forward and rs emission have been observed as separate components. 
In these, we fit the dominant component (the one with the higher $EM$, which is normally better constrained observationally)
 and then check if the other predicted component is consistent with the other observed component.
%For most cases, the ejecta composition is not measured by the X-ray spectrum, 
%thus we take it to be that for Type Ia or massive star, as given above. If it is measured, we use the measured composition.

The model here has more parameters (7) than the number of observed parameters of the SNR (3 if one pair of $EM$ and $T$ is observed
or 5 is two pairs are observed).
Thus we make further simplifications.
$T_{ISM}$ is set to 100K, which doesn't affect the evolution unless the SNR is very old;
and we apply both $s=0$ and $s=2$ models, effectively reducing the number of model parameters to 5.
Because s=2 models have much faster growth of the fs radius with time, the two cases give very different parameters, 
such as age and explosion energy. 
We normally choose which of the two models yields results most consistent with any other observational data.
%The cases of 3 and 5  observed parameters are treated differently:
\begin{itemize}
\item For 3 observed parameters, we normally fix  n to be 7 and fix  $M_{ej}$, leaving 3 model parameters: age, energy and $n_0$ (s=0) or $\rho_s$ (s=2).
For cases where the SN type is core-collapse (hereafter CC) we take $M_{ej}$=5$M_{\odot}$, and
for cases where the SN is Type Ia or unknown, we take $M_{ej}$=1.4$M_{\odot}$.
\item For 5 observed parameters, we first fit the dominant emission component. Then we adjust n and $M_{ej}$
to approximately match the  $EM$ and $T_{EM}$ of the other component.
In effect, there are 5 model parameters : n, $M_{ej}$, age, energy and $n_0$ (s=0) or $\rho_s$ (s=2).
\end{itemize}

In summary, some SNRs have 3 inputs: fs radius, and emission measure and temperature of either fs or rs.  
This results in 3 outputs: age, explosion energy, and either ISM density (for the $s=0$ case) 
or CSM density parameter (for the $s=2$ case).
Some SNRs have 2 additional inputs ($EM$ and $T$ for both fs and rs material), 
which allows inference of additional constraints on n and $M_{ej}$, resulting in 5 model outputs.  

It often not known whether the main emission component is from fs or rs material. 
In such cases, we apply one model assuming emission from the fs, and one assuming emission from the rs. 
Because the model calculates $EM$ and $kT$ for both fs and rs, we can determine which model is more 
consistent with the data. In some cases, both models are consistent. 
For those cases, other information (e.g. kinematic expansion age) is used to decide which model is more likely correct.
 
\section{The SNR Sample}\label{sec:sample}

For the inner Galaxy, a significant number of SNRs (30) have had their distances  measured or re-measured using HI absorption and 
$^{13}$CO emission data.
Details of distance determinations are given in \citet{2017SR1}, \citet{2018SR4}, \citet{2017SR2} and \citet{2018SR3}.
Of these, 15 have existing X-ray spectral observations which yield EM and temperature. 
These SNRs include 2 cases with new distances (no previous values), 
9 cases with changed distances and 4 cases with distances confirmed to be the same as from previous studies.

This set of 15 SNRs comprises our sample for modelling. 
For each SNR, the distance comes from the HI and CO studies and the EM and EM-weighted temperature comes from the X-ray studies. 
We examined both the VGPS radio images (available for all SNRs) and the Chandra X-ray images (available for some) 
to measure angular semi-major and semi-minor axes of the outer shock. 
The average of major and minor axes is used to estimate the average outer shock radius as input to the spherical SNR model. 
The observational input parameters are summarized in Table~\ref{tab:TBLobserved}.

\subsection{Models for Individual SNRs in the Sample}\label{subsec:indSNR}

\subsubsection{G18.1-0.1}
G18.1-0.1 was noted as a probable SNR by \cite{1986Ode} and confirmed as an SNR by \citet{2006Brogan}. 
The distance was determined by \citet{2014Leahy} as 5.6 kpc. 
That same work analyzed the Chandra X-ray spectrum of G18.1-0.1.  
Application of a Sedov model gave an age of $\sim$5 kyr and a low explosion energy. 
The improved distance analysis of \citet{2018SR4} gave a distance of 6.4$\pm$0.2 kpc, which is used here.

We use the $EM$ and $kT$ from the apec model in Table 2 of \citet{2014Leahy}\footnote{For all of the SNRs in this section we 
convert the $EM$ from XSPEC units (or other units given in the paper) to units of $10^{58}$cm$^{-3}$ using the best distance
for that SNR. If in XSPEC units, the conversion factor is $10^{14}~4\pi~d^2$ with d the distance in cm. }.
Our model for uniform ISM (s=0, n=7) for this SNR yields an age of 5400 yr and low explosion energy of 0.19$\times10^{51}$ erg.
For this model, $t_{rev}=3600$ yr and  $t_{pds}=9800$ yr so the SNR is in the adiabatic phase.
The predicted EM of the rs ejecta is 0.032$\times10^{58}$cm$^{-3}$, much fainter than for the fs.
The stellar wind SNR model does not yield a reasonable age (500 yr), so is not used.

\subsubsection{G21.5-0.9}

G21.5-0.9 consists of a bright central pulsar wind nebula and a faint surrounding shell, thus is CC type.
The spin-down age of the pulsar PSR J1833-1034 in G21.5-09 is 4.9 kyr \citep{2005Gupta}. However
an expansion speed measurement of the SNR yields an age of  870$^{+200}_{-150}$ yr \citet{2008Biet}, making it very young.
The  distance to this SNR is given by \citet{2018SR4}.
The Chandra X-ray image and spectrum was analyzed by \citet{2010Matheson}. They analyze the extended emission
using a two component model consisting of a power-law plus pshock (their Table 4). 
The power law represents the contribution of synchrotron to the
X-ray emission and the pshock represents the gas shocked by the SNR. 
We use $EM$ and $kT$ from the pshock component from the
southern halo \citep{2010Matheson}, and multiply the $EM$ by a factor of two to account for the emission from the northern half of the SNR. 

We apply a SNR model for uniform ISM (s=0) with n=7, and 5 $M_{\odot}$ ejecta assuming the fs is providing the X-ray emission.
This yields an age of 2450 yr and low explosion energy of 0.019$\times10^{51}$ erg.
For this model, $t_{cor}=3600$ yr so the SNR is in the early ejecta-dominated phase. No consistent models are found for later stages.
The predicted EM of the rs is about half of that for the fs and of similar temperature. 
It would be difficult to distinguish from the fs emission, and thus the model is consistent with a single observed emission component.

Next we apply s=0, rs emission models. 
The s=0, n=7 model yields an age of 1240 yr, explosion energy of 0.085$\times10^{51}$ erg and $t_{cor}=1550$ yr. 
The predicted EM of the fs is nearly the same as for the rs, and the fs has a temperature of 1.1 keV.
Alternately, we apply the s=0, rs emission model but with n=9. 
This yields an age of 470 yr, explosion energy of 0.56$\times10^{51}$ erg and $t_{cor}=540$ yr. 
For n=9, the predicted EM of the fs ejecta is one-third of that for the rs and the fs has a temperature of 3.6 keV.
The fs component would be difficult to distinguish from the bright PWN emission and the rs emission, so is consistent with observations.
Because the s=0 n=9  rs emission model is consistent with the expansion age, we prefer that model over the s=0 fs model and adopt it.\footnote{The age decreases as n increases for the rs model. n=8 gives an age of 775 yr, n=10 gives an age of 280 yr. 
This steep dependence of age on n occurs for the rs emission models because the rs EM is sensitive to n.} 

The stellar wind SNR fs and rs models yield unreasonable ages ($\lesssim$200 yr), so are not used.

\subsubsection{G21.8-0.6 (also known as Kes 69)}

G21.8-0.6  was observed in X-rays with XMM-Newton by \citet{2013Seo}. 
The VLA radio image and kinematic distance to the SNR are given by \citet{2008Tian}.
The SNR consists of an incomplete radio shell and center-filled X-ray emission.
The distance has been updated by \citet{2018SR4} to 5.6$\pm$0.2 kpc. 

The  $EM$, $kT$ and uncertainties are from Section 3 and Fig. 3 of \citet{2013Seo}.
A consistent SNR model has s=0, n=7, 1.4 $M_{\odot}$ fs emission. 
This is our adopted model.
The SNR is 9700 yr old, in adiabatic stage with $t_{rev}=6200$ yr and $t_{pds}=45000$ yr.
The predicted emission measure of the rs ejecta is 0.003$\times10^{58}$ cm$^{-3}$,  much fainter than for the fs.

\subsubsection{G27.4+0.0  (also known as Kes 73)}

G27.4+0.0 is a shell type SNR with an anomalous X-ray pulsar at its center \citep{1994Helf,1997Vas}. %,  \citep{1997Vas}. 
The distance of 8.5 kpc was determined by \citet{2008Tian}.
It was observed in X-rays with Chandra and XMM-Newton by \citet{2014Kumar}. The X-ray spectrum yielded two components:
one with low EM and high kT and a second with high EM and low kT. The first component had solar abundances, so is 
associated with shocked ISM, and the second had enhanced O, Si and S abundances, so is associated with the ejecta.
\citet{2017Bork} measured the expansion velocity of the SNR shell as 1100 km/s, assuming distance of 8.5 kpc, and derived
a Sedov age of 1800 yr.
\citet{2018SR4} revised the distance to 5.8$\pm$0.3kpc, using a better rotation curve model for the inner Galaxy.

The  unabsorbed flux and $kT$ are from \citet{2014Kumar} Table 3 whole ``SNR'' column values. 
Their unabsorbed flux was converted to $EM$ using XSPEC with the spectral parameters fixed to their values.
We applied a Sedov model using the new distance (and thus radius) obtaining an age of 1600 yr (Table~\ref{tab:TBLST}),
similar to that obtained by \citet{2017Bork}. However the ST model ignores the effect of ejecta mass, which can considerably affect
age and explosion energy estimates.

For this SNR  \citep{2014Kumar}, we know the high kT, low EM component is fs emission, and the low kT, high EM component is rs emission.
We apply the s=0, n=7, 5 $M_{\odot}$ ejecta, fs emission model to obtain an age of 2160 yr and explosion energy of 0.33$\times10^{51}$ erg.
 For this model, $t_{cor}=980$ yr and $t_{rev}=7300$ yr,  so the SNR is in the transition between ejecta-dominated and adiabatic phases. 
However, the predicted EM of the rs is lower (by a factor of $\simeq$0.06) than that for the observed rs component,
so is not consistent with observations.
For the s=0, n=12, fs emission model, the age reduces to 2020 yr and the explosion energy increases to 0.34$\times10^{51}$ erg.
This model has rs EM increased to 49.2$\times10^{58}$cm$^{-3}$, which now agrees with the rs EM observed, 
and rs temperature of 0.77 keV.
Increasing the ejected mass from 5 to 10 M$_{\odot}$ for the s=0, n=12, fs emission model, yields
an age increase to 2490 yr,  explosion energy decrease to 0.33$\times10^{51}$ erg,
rs EM increase to 52.7$\times10^{58}$ cm$^{-3}$ and rs temperature decrease to 0.51 keV. 
These values are consistent with the observed values, so this last model is the adopted model.

The stellar wind SNR models for n=7 to 12 yield very low ages (150-200 yr) and very high explosion energies 
($\simeq30-60\times10^{51}$ erg), so are not used.

\subsubsection{G28.6-0.1}

G28.6-0.1 was confirmed as an SNR by \citet{2003Ueno} using Chandra observations of the region containing the SNR candidate
AX J1843.8-0352, identified by \citet{2001Bamba}. 
The distance to this SNR was obtained by \citet{2017SR2}. 
The unabsorbed flux and $kT$ are from the Table 2 NEI model of \citet{2003Ueno}. 
Their unabsorbed flux was converted to $EM$ using XSPEC with the spectral parameters fixed to their values.

We apply the s=0, n=7, 1.4 $M_{\odot}$ ejecta, fs emission model. This is the adopted model.
This SNR is faint and of large size. It is in a low density region, is relatively old (age $\sim15,000$ yr) and in the the adiabatic phase 
($t_{pds}=101,000$ yr). The model predicted rs emission is very faint (EM of $10^{55}$cm$^{-3}$).

The stellar wind s=2 SNR models yield low age ($\sim1300$ yr) and very high explosion energy $23\times10^{51}$ erg) so are not used.

%We consider a model with the high EM and low kT component as the fs ejecta. 
%This yields an age of 1780 yr and explosion energy of 0.58$\times10^{51}$ erg.
% For this model, $t_{cor}=650$yr and $t_{rev}=4800$yr  so the SNR is in the transition between ejecta-dominated and adiabatic phases. 
%The predicted reverse shock EM is $simeq$5\% of that of the
%and the predicted forward shock temperature is $\simeq$0.6 keV, so it would have been observed.
%The SNR is 2150 yr old, in the transition from ejecta-dominated stage to adiabatic stage with $t_{cor}=990$yr and $t_{rev}=7300$yr.
%so that it would be difficult to distinguish from the forward shock emission.

\subsubsection{G29.7-0.3  (also known as Kes 75)}

G29.7-0.3  was identified as a composite SNR by \citet{1984Bec}. \citet{2009Su} found molecular clouds interacting
with the SNR and analyzed the Chandra X-ray spectra of the diffuse shell, away from the central pulsar wind nebula.
Their Fig. 8 shows the diffuse X-ray emission from the SNR consists of the SE and SW regions.
We use the mean $kT$ from the two spectral fits in their Table 3, and sum the normalizations to determine the $EM$ of the SNR.
The first reliable distance of 5.1-7.5 kpc to G29.7-0.3 was given by \citet{2008LeahyA}, which was updated by \citet{2018SR4} to 5.6 kpc. 

An s=0 n=7, 5 $M_{\odot}$ ejecta, fs emission model yields an age of 2590 yr and explosion energy of  0.057$\times10^{51}$ erg.
The SNR is in the ED stage, with age just less than $t_{cor}$=2750 yr. The predicted rs emission has EM of 2/3 that of the 
fs and temperature of 0.33  keV. 

An  s=0 n=7 rs emission model yields an age of 1650 yr and explosion energy of 
 0.16$\times10^{51}$ erg. It is in the transition from ED to adiabatic stage with $t_{cor}$=1520 yr and $t_{rev}$=11300 yr.
 The predicted fs emission has EM 1.6 times that of the rs and temperature of 1.4 keV, thus is not consistent.
 However an s=0 n=9 rs emission model yields age of 890 yr and explosion energy of  0.46$\times10^{51}$ erg.
 The predicted fs emission has EM 0.37 times  that of the rs and temperature of 2.7 keV, so is
consistent with the observations.

Stellar wind s=2 SNR models yields low age ($\sim170$ yr) and  high explosion energy $8\times10^{51}$ erg) so are not used.
Because the measured abundances are indicative of rs emission we adopt the s=0, n=9 rs model and use it in Table~\ref{tab:TBLfullmodel}.
%In reality, for this SNR probably both the reverse and forward shocked material are observed, but a better X-ray spectrum is needed to separate their contributions.

\subsubsection{G31.9+0.0 (also known as 3C391)}

G31.9+0.0 is a mixed morphology SNR \citep{1993Reyn} and is interacting with a molecular cloud \citep{1998Wilner}. 
The distance to G31.9+0.0 is 7.1$\pm$0.4 kpc \citep{2017SR1}. 
\citet{2014Sato} carried out a high sensitivity X-ray study of this SNR using Suzaku observations. 
They determined the X-ray spectral parameters of a two-component plasma in the SNR: a collision-ionization equlibrium  (CIE) componnent and
a recombining plasma (RP) component. The abundance patterns of the RP are consistent with a core collapse SN. 
The $kT$ and $VEM$ are from their recombining plasma model (CIE plus NEIJ) Table 3 (third and fourth columns). 
The $VEM$ was converted to $EM$ using the new SNR distance.
The CIE+CIE $kT$ and $VEM$ values are very similar to those for the CIE plus NEIJ model.
%The $\chi^2$ of the two models are also similar, with the recombining plasma model slightly better.

Initially we use an ejecta mass of 5 $M_{\odot}$ in our models. 
An s=0, n=9, fs emission model for the CIE component yields an SN explosion energy of  0.39$\times10^{51}$ erg, 
age of 8740 yr and ISM density of 4.53 cm$^{-3}$.
This model gives rs kT=0.38 keV and EM=0.61$\times10^{58}$cm$^{-3}$. The rs EM is within the observed RP-EM uncertainty range
but on the low side and the rs kT is close but lower than the observed RP value.
 The n=7 and n=12 models give almost the same results as the n=9 model for age, explosion energy and EM, kT for the rs. 

The EM of the rs for the s=0, n=9, fs model can be increased by increasing the ejecta mass. 
For example increasing the ejecta mass to 15 M$_{\odot}$ increases the rs EM to 1.9$\times10^{58}$cm$^{-3}$, but
lowers the rs kT to 0.26 keV. As a compromise to fitting both EM and kT for the rs we choose the s=0, n=9, 
10 M$_{\odot}$ model to present in Table~\ref{tab:TBLfullmodel}. 

SNR in stellar wind (s=2) models yield much higher EM and lower kT for the rs component compared to the observed RP values,
so are not used.

Because this is a mixed-morphology SNR, we apply the WL \citep{1991WL} model. Our WL model includes Coulomb electron-ion equilibration.
We choose an intermediate cloud evaporation parameter ($C/\tau=2$) which has flat interior density and temperature profiles (Fig. 2 of  \cite{1991WL}),
similar to those seen for mixed-morphology SNRs. The WL model does not include any ejecta, but if the SNR is old enough that the emission
is dominated by ISM and evaporated cloud emission, the model is a good approximation. The results of the WL model for G31.9+0.0 are 
given in  Table~\ref{tab:TBLfullmodel}. It has nearly the same age and explosion energy as the above adopted model. 
The WL model has a lower ISM density, as expected, because evaporating clouds contribute to the observed $EM$.
% This model also gives a $t_{PDS}$ of 10000 yr, so that the forward shock is not radiative yet. 
%Our current inverse SNR model does not include a treatment of shock cooling by radiative losses. 
%However we can estimate the effects of cooling by comparing non-radiative andradiative SNR forward evolution models as described by \citet{2017LeahyWilliams}.

\subsubsection{G32.8-0.1 (also known as Kes 78)}

G32.8-0.1 is an SNR interacting with a molecular cloud \citep{2011Zhou}.  
The distance was revised to 4.4 kpc by \citet{2018SR4}. 
The most recent X-ray study of this SNR is presented by \citet{2016Bamba}, using Suzaku data. The SNR diffuse 
emission has an X-ray spectum which was equally well fit with two different two-component models: low (0.63 keV) and high 
(3.4 keV) temperature VNEI components, or a low temperature VNEI component plus a power-law component. 
The latter model is preferred by \citet{2016Bamba}, 
and we use $EM$ (converted to the new distance) and $kT$ from that model (their Table 3).

Because there is no clear evidence that this is core-collapse SN, we use an ejected mass of 1.4 M$_{\odot}$ and n=7 for modelling. 
We ran s=0 fs models for different stages of evolution and 
found a consistent model if the SNR has an age between $t_{cor}$ and $t_{rev}$. 
The predicted EM of the rs component is 1.4$\times10^{56}$ cm$^{-3}$ in agreement with the EM inferred for the second component of
the two-component VNEI model. However the predicted kT for the rs component is a factor of $\sim$10 below the 
kT of the second VNEI component. 

An s=2, n=7 model yields an age of 630 yr and explosion energy of 5.6$\times10^{51}$ erg, with rs EM  1.0$\times10^{56}$ cm$^{-3}$.

Neither the s=0 nor s=2 models can give a 3-4 keV rs, so we agree with  
\citet{2016Bamba} that the second component appears more likely to be non-thermal.
The low ISM density inferred from the SNR models results from the large size and low EM. 
This is consistent with G32.8-0.1 interacting with
a molecular cloud if the cloud interaction has taken place fairly recently compared to the age of the SNR. 
The adopted model is the s=0, n=7, 1.4 M$_{\odot}$ fs emission model.

\subsubsection{G33.6+0.1  (also known as Kes 79)}

G33.6+0.1 is a mixed morphology SNR \citep{1998Rho} and a GeV gamma-ray emitter \citep{2014Auch} interacting
with a molecular cloud. The distance was revised to 3.5 kpc  \citep{2018SR4}, significantly closer than previously estimated.
The X-ray spectrum was characterized using Suzaku observations \citep{2016Sato}. Two emission components
are found, an NEI plasma associated with the rs and a CIE plasma associated with the fs.
We use the $EM$ and $kT$ from Table 3 Model B with $EM$ converted to cm$^{-3}$ using the new distance.
Based on the abundance pattern in the ejecta, \citet{2016Sato} suggest a high mass progenitor of $\sim40$M$_{\odot}$.

We initially used an ejecta mass of $20$ M$_{\odot}$ in our models. 
Consistent s=0 models are all in the evolutionary stage between $t_{cor}$ and $t_{rev}$.
The age depends on the ejecta mass with higher ejecta mass yielding older models. E.g. for n=9 models, the age is 8300 yr for  $5$ M$_{\odot}$
ejecta and 10300 yr for  $20$ M$_{\odot}$ ejecta. The models produce too high EM for the rs and too low kT compared
to measured values, unless the ejecta mass is low, $\simeq$2 M$_{\odot}$.

Next we ran s=2 models to fit the fs emission. For the s=2 models, a given n has a fixed ratio of EM of the fs to EM
of the rs, independent of ejecta mass. This is a result of the  self-similar nature of  s=2 models. 
Thus the observed EM ratio requires n of 6 to 7 with higher n ruled out. For s=2 models, the age is also independent of
ejecta mass. 

The s=0, n=6 and n=7 models give ages of 3100 yr and 780 yr, respectively. For an ejected mass of 20 M$_{\odot}$,
these models have explosion energies of 0.48$\times10^{51}$ erg and 2.93$\times10^{51}$ erg, respectively.
The s=2, n=7 model has rs EM and kT close to the observed values. 
In  Table~\ref{tab:TBLfullmodel} below we give the parameters for this model.
We note that the s=2, n=7 model with 40 M$_{\odot}$ ejecta has the same age and rs EM and kT as the 20 M$_{\odot}$, 
but an explosion energy of 8.64$\times10^{51}$ erg.

\subsubsection{G34.7-0.4  (also known as W44)}

G34.7-0.4 is a mixed morphology SNR with associated CO emission \citep{2004Seta,2005Reach}.  
ASCA spectra determined that the plasma was nearly in CIE \citep{2005Kawa}. \citet{2012Uchida} reported the analysis of Suzaku
X-ray spectra of W44 demonstrating the presence of a recombining plasma. The X-ray image in 1-2 keV and 2-5 keV bands shows the
centre filled nature of the thermal plasma X-ray emission from W44, while the 5-7 keV image shows the hard X-rays are displaced and
likely from a different mechanism \citep{2012Uchida}.
The abundances of the thermal plasma are mildly enhanced, with Si, S, Ar and Ca all about twice solar. This indicates that
the emission is from a mixture of ISM and ejecta. Although the deduced amount of ISM and ejecta contributions depend on rather
uncertain ejecta abundances, the low enhancement indicates the emission is likely dominated by ISM emission. 
The acceptable plasma model from
\citet{2012Uchida} is the nonequilibrium ionization jump (NEIJ) model. This model describes a plasma which is in CIE at temperature kT$_{e1}$,
then has electron temperature dropped suddenly to kT$_{e2}$ at some time in the past. The parameter n$_{e}$t is the product of
electron density and time since the temperature was dropped. The cause of the drop is plausibly adiabatic expansion, which
would be caused if the blast wave was initially in dense circumstellar environment then propagated into the low density interstellar medium.
An estimate of the time since the rapid cooling is $\sim$ 15000 to 20000 yr  \citep{2012Uchida}.

We use the combination of regions A, B and C (see their Fig. 1) to represent the emission from the whole SNR.  $kT$ is taken as the average
of the 3 values of $kT_{e2}$ from their Table 2, and $EM$ is the sum of their $EM$s, converted to the new distance to the SNR. 
 The distance to this SNR was updated by \citet{2018SR4}.

Since most of the SNR blast-wave evolution has occured after the sudden cooling, the appropriate temperature to use for our modelling
is  kT$_{e2}$.  We use the sum of the observed emission measures of the different regions to represent the whole SNR emission.
We use 5 M$_{\odot}$ ejecta, and have verified that for an old SNR, the ejecta mass make very little difference to the model. 

The s=0, n=7, fs emission model yields an SN explosion energy of  2.23$\times10^{51}$ erg, age of 9100 yr,
an ISM density of 1.52 cm$^{-3}$
and significant rs emission (EM=11$\times10^{58}$cm$^{-3}$), about 7.5\% of the EM of the fs.

An s=2, n=7, fs emission model has a low age ($\sim$1400 yr) and high energy (10.2$\times10^{51}$ erg). 
The model predicts rs EM and kT of 4.6$\times10^{59}$cm$^{-3}$ and 0.47 keV. 
This is roughly consistent with it being observed together with the fs emission which has kT = 0.49 keV. 
The only issues with the s=2 model are the low age and high energy.
Thus we adopt the s=0, n=7 model.

Because this is a mixed-morphology SNR, we apply the WL model with $C/\tau=2$.  The results of the WL model for G34.7-0.4 are 
given in   Table~\ref{tab:TBLfullmodel}. It has nearly the same age and explosion energy as the above adopted model. 
The WL model has a lower ISM density, as expected, because evaporating clouds contribute to the observed $EM$.

\subsubsection{G39.2-0.3  (also known as 3C396)}

G39.2-0.3 is a composite SNR with central pulsar wind nebula and outer shell seen in radio \citep{1990Patnaik,1993Anderson} and X-ray \citep{1987Bec}.
Shocked molecular gas has been identified with 3C396 \citep{2009Hewitt}. The updated distance to G39.2-0.3 is 8.5 kpc \citep{2018SR4}.
\citet{2011Su} carried out X-ray spectroscopy of the  thermal X-ray emission from this SNR, detecting two components interpreted as the shocked ISM and
the shocked ejecta. 
Their N and S regions (see their Fig. 9 and Table 2) have enhanced abundances and high $kT$, whereas the E, W and SW regions have near
solar abundances and low $kT$. We agree with their interpretation and take the shocked ejecta emission to be N+S  ($kT$ and $EM$ from N+S fit).
We take the shocked ISM emission to be E+W+SW, and use the $kT$ value of the E region, because it is the same as the mean of the E,W and SW values.
We sum the $EM$s of the E, W and SW regions and convert to cm$^{-3}$ using the new distance to this SNR.

We apply our SNR models to this CC--type SNR. We model the shocked ISM component first using an s=0, n=9,  
5 M$_{\odot}$ ejecta, fs emission model. 
The only consistent solutions are when the SNR is in the transition stage, between $t_{cor}$ and $t_{rev}$.
The model yields a predicted rs EM of 4.9$\times10^{58}$ cm$^{-3}$ and rs kT of 0.59 keV.
An n=7 fs model yields a  rs EM of 1.1$\times10^{58}$ cm$^{-3}$ and kT of 0.68 keV, close to the observed values. 
An n=6 fs model gives higher rs kT of  0.68 keV, but lower rs EM of 3.2$\times10^{57}$ cm$^{-3}$. 
The n=7 model as closest to the observations.

A s=2, n=7, fs emission model yields low SNR age (470 yr), high explosion energy  (23$\times10^{51}$ erg) and rs EM and kT very
similar to the s=0, n=7 model. Because of the high energy and low age of s=2 models, the s=0, n=7, fs emission model is adopted.
 
\subsubsection{G41.1-0.3  (also known as 3C397)}

G41.1-0.3 is a composite SNR and is seen as a shell in radio observations \citep{1993Anderson} and as centrally concentrated in 
X-rays \citep{1998Rho}.  
\citet{2005Safi} presents the Chandra X-ray study of this SNR. 
The spectral fitting for the diffuse emission of the bright and large Eastern and Western Lobe regions (see Fig. 6 and Table 3) was
used to represent the spectrum of the whole SNR. $kT$ was taken from the Western Lobe, because that is equal to the mean of the
two Lobes and better represents the uncertainty. $EM$ was taken as the sum of Eastern and Western lobes, multiplied by 2 to
account for the missing diffuse emission (see their Fig. 6), then converted to cm$^{-3}$ using the new distance of the SNR.
Based on the ejecta abundances, G41.1-0.3 was  identified to be a Type Ia remnant  with an energetic explosion \citep{yamaguchi2015}.
A new distance was obtained and modelling of this SNR was carried out by \citet{2016Leahy}.
That gave a low age and confirmed an energetic explosion. The updated distance to G41.1-0.3 is 8.5 kpc \citep{2018SR4}.

Because it is type Ia, we apply s=0, 1.4 M$_{\odot}$ ejecta models using the type Ia ejecta compositions given in Section 2.2.
We find consistent models for ages $>t_{rev}$.
The model with n=7 predicts  EM  of the rs to be 0.19$\times10^{58}$ cm$^{-3}$ closer to the observed values than for other values of n.
The model gives kT  of the rs as 6.0 keV. This could be a result of our assumption of uniformly mixed ejecta composition.
For example, if we instead use typical CC ejecta composition which has lower mean ion mass, the rs temperature is 2 keV.
Fitting the ejecta composition would likely allow fitting the rs EM and T.  However that is a complex procedure beyond
the scope of the current work.

A stellar wind SNR model (s=2, n=7) yields a lower age (1300 yr) and explosion energy of  0.95$\times10^{51}$ erg.
 The predicted EM and kT of the rs are 3.0$\times10^{57}$ cm$^{-3}$ and 0.43 keV. In this case the EM is close to the
observed value but the temperature is low. Adjusting the ejecta composition could likely give a temperature closer to the observed value. 
The s=2, n=7 model  has explosion energy consistent with that deduced by  \citep{yamaguchi2015}.
Thus the adopted model is the s=2, n=7 one. %and are listed in Table~\ref{tab:TBLfullmodel}.

\subsubsection{G43.3-0.2  (also known as W49B)}

G43.3-0.2  is a mixed morphology SNR \citep{1998Rho}. 
 The X-ray spectrum was shown to consist of a solar abundance component and a heavy element enriched component \citep{2000Hwang}, and noted as a probable type Ia SNR.
G43.3-0.2 was discovered to have a recombining plasma by \citet{2005Kawa}.
With a long Suzaku observation, the integrated X-ray spectrum of G43.3 was analyzed by \citet{2009Ozawa}. This confirmed  the presence of a recombining plasma.
Because this analysis gives the spatially integrated X-ray spectrum, we use their parameters for $kT$ and $EM$ for the whole SNR found
using the CIE model (see their Table 1 and Section 4.4), with their $VEM$ converted to cm$^{-3}$ using the newest distance to the SNR . 
\citet{2013Lopez} illlustrate its unusual X-ray morphology (their Fig.3), with an iron-rich bar. 
They analyze Chandra X-ray spectra of $\sim$700 small regions, 
%finding that they can be fit with a one-component enriched abundance plasma. The conclusion of that study was 
and conclude that G43.3-0.2 likely originates in a jet-driven core-collapse explosion, rather than a type Ia.
Because of the conflicting evidence for type Ia or CC, we take the SN type of G43.3-0.2 as unknown.
Further analysis by \citet{2013LopezB} found that the origin of the recombining plasma was likely adiabatic expansion of the hot plasma.
The updated distance to G43.3-0.2 is 11.3 kpc \citep{2018SR4}.

We apply models with ejecta masses of 1.4 M$_{\odot}$ for a type Ia explosion and 5 or 10 M$_{\odot}$ for a CC explosion.
The observed abundance of Fe and Ni are $\simeq$ 5 times solar, but the large shock radius and large $EM$ imply the emission is dominated by swept up ISM.
The  iron abundance is elevated (5 times solar) but it is not clear if it is dominated by ISM or ejecta emission.
The swept up mass for W49B is estimated as $(4\pi/3)~n_0~R^3~\simeq~ 80~M_\odot$, using an estimated $n_0=1$ cm$^{-3}$.
An iron rich ejecta of $\sim1~M_\odot$, would yield the observed abundance enhancement, so the emission is  mostly from the ISM.
Thus we apply fs models.

The s=0, n=7, 9 or 12, and 1.4 or 5 M$_{\odot}$ ejecta models all have age larger than $t_{rev}$. 
The age is 3240-3250 yr for the 1.4  M$_{\odot}$ ejecta cases, and  3400-3430 yr for the 5  M$_{\odot}$ ejecta cases.
The predicted rs $EM$ is small enough (1\% of fs EM for 5 M$_{\odot}$, and 5\% of fs EM for 1.4 M$_{\odot}$) 
to be consistent with the observed X-ray spectra, which is dominated by fs EM.
The high predicted rs $kT$ ($\simeq$2.8 keV for 5 M$_{\odot}$ ejecta, $\simeq$7.8 keV for 1.4 M$_{\odot}$ ejecta) might be related to the
observed recombining plasma in this SNR (see Section 4.3 below). 

The s=0,  n=7, 9 or 12 with 10 M$_{\odot}$ ejecta models evolve more slowly, so they are in the stage $t_{cor}<t<t_{rev}$. 
However these models all have the rs emission with EM $\sim$5 times larger than the fs, so are not consistent with observations.
%For the stellar wind SNR models (s=2), we ran different n values for ejecta masses of  1.4M$_{\odot}$ and 10M$_{\odot}$. 

For a SNR in a stellar wind (s=2) with 1.4 M$_{\odot}$ ejecta, the n=7 model has the fs emission brighter than the rs, 
but higher n models have the rs brighter so that they are inconsistent with the observations.
However, for all n, they require high explosion energies (0.6 to 5$\times10^{53}$ erg) and yield low ages (290 to 690 yr).
For a SNR in a stellar wind (s=2) with 5 or 10  M$_{\odot}$ ejecta the ages and rs $EM$s are similar to the 1.4  M$_{\odot}$ ejecta, but the 
explosion energies are $\simeq$3 (for 5  M$_{\odot}$ case) and  $\simeq$6 (for 10 M$_{\odot}$ case) times larger.
Thus the s=2 models are not realistic. 

Only the  s=0, n=7, 9 or 12, and 1.4 or 5 M$_{\odot}$ ejecta models are consistent with observations. 
We have a preference for the s=0, n=7, 1.4  M$_{\odot}$ model because of the elevated iron abundance but not intermediate mass elements. 
This is characteristic of type Ia explosions, and we adopt that model.

Because this is a mixed-morphology SNR, we apply the WL model with $C/\tau=2$.  The results of the WL model for G43.3-0.2 are 
given in   Table~\ref{tab:TBLfullmodel}. It has nearly the same age and explosion energy as the above adopted model. 
The WL model has a lower ISM density, as expected, because evaporating clouds contribute to the observed $EM$.

\subsubsection{G49.2-0.7  (also known as W51C)}

G49.2-0.7 is  part of the W51 complex containing prominent HII regions W51A and W51B.
and has been detected in $\gamma$-rays \citep{2009Abdo}. 
The HI absorption spectrum was analyzed by \cite{2013Tian}, who found that G49.2-0.7 is this side of the tangent point and not associated with high velocity HI.
The radio emission of the SNR was modelled using magneto-hydrodynamic simulations by \cite{2017Zhang}, yielding a new angular diameter and giving
an estimate of the ejecta mass as 11M$_{\odot}$. 
 Here we use an angular radius for the SNR of 21 arcminutes, based on comparison of the radio image
with the magneto-hydrodynamic simulations. The SNR distance was updated to 5.6 kpc by \cite{2018SR4}.

G49.2-0.7 has been studied a number of times in X-rays, including an analysis of Chandra observations by  \citet{2005Koo},
\citet{2013Hanabata} analyze Suzaku observations of the diffuse thermal X-ray emission of G49.1-0.7 and find it is best fit with an NEI model. There is
an additional hard X-ray component associated with the overlap of the SNR and adacent molecular clouds, which is not well understood but could
be associated with synchrotron X-rays from TeV electrons. 
\citet{2014Sasaki} analyze XMM-Newton observations of the SNR and final similar results in their spectral analysis but suggest that the hard component
originates in a pulsar wind nebula. They note that their extended region 1 (south-eastern half of the SNR) has solar abundances but their region 2 (the
western half) has enhanced Ne and Mg abundances, and infer a progenitor mass of $\gtrsim$20M$_{\odot}$.  %no emission measure given

We model the thermal component here
and use the spectral parameters given in Table 2 of \citet{2013Hanabata} for the NEI model fits for Regions 1 and 2 combined.
The $kT$ for Regions 1 and 2 are nearly identical, so we use that for Region 1. We sum the $EM$ values of Regions 1 and 2, which contain
most of the 0.5-2.5 keV flux from the SNR (see their Fig. 2), then convert the sum to cm$^{-3}$ using the newest distance to this SNR.

We first apply s=0, 10 M$_{\odot}$ ejecta, fs emission models. For for different n, these are all in the stage with $t_{core}<t<t_{rev}$.
The n=7 model produces fs emission much higher than for the rs, so is not consistent with observation of enhanced Ne and Mg abundances.
n=9 to 12 models produce similar $EM$ and $kT$ for fs and rs. 

We next apply s=0, 10 M$_{\odot}$ ejecta, rs emission models. %For different n, these are all in the stage with $t_{core}<t<t_{rev}$.
These produce results consistent with the fs emission models. I.e. the n=7 model produces fs emission much higher than for the rs, 
so is not consistent with observations. n=9 to 12 models produce similar $EM$ and $kT$ for fs and rs. 
%We choose to use the fs emission models in preference to the rs models because the uncertainties of the fs model are smaller.

It is likely that the observed X-ray emission has contributions from both forward and rs emission.
Thus the models support the suggestion by \citet{2014Sasaki} that their region 1 is shocked ISM and their region 2 is shocked ejecta.
The best matching fs emission model with 10 M$_{\odot}$ ejecta has n=9. 
For this model the age is 16000 yr, the explosion energy is 0.76$\times10^{51}$ erg and ISM density is 0.021 cm$^{-3}$.
The fs and rs temperatures are nearly the same (0.70 and 0.61 keV) and the fs and rs each contribute half of the observed
emission measure. The s=0, n=9, fs emission model is the adopted model.

We calculated an s=0, n=9, fs emission model with 20 M$_{\odot}$ ejecta because of the high suggested progenitor mass. 
The age is 18000 yr, the explosion energy is 0.73$\times10^{51}$ erg and the ISM density is 0.020 cm$^{-3}$, not much different than
the adopted model. 
For all of our models the ISM density is low. 
G49.1-0.7 has expanded into a low density medium prior to colliding with the molecular cloud on its western side (e.g. see Fig. 1 of \citealt{2005Koo}).

Stellar wind SNR models (s=2) require large energies ($\sim$ 2$\times10^{52}$ erg) and low ages ($\sim$3000 yr), so are considered unlikely.

\subsubsection{G54.1-0.3}

G54.1-0.3 is a Crab-like SNR \citep{2002Lu} with the central region dominated by non-thermal X-ray emission. 
Based on the pulsar spin-down age of 2900 yr, the SNR age is estimated at 1500-6000 yr \citep{2002Camilo}.
\citet{2010Bocchino} use XMM-Newton and Suzaku observations of G54.1-0.3 to detect the thermal emission in the faint outer regions of the SNR. 
They corrected for scattering of the bright central PWN emission into the outer shell region of the SNR.  
We use the spectral parameters from their fit to the SNR shell spectrum, in their Table 2, and use XSPEC to 
convert their unabsorbed flux for the shell into XSPEC norm, then into $EM$ using the new distance to the SNR.
 The SNR distance was first determined by \citet{2008Leahy}, then updated to 4.9 kpc by \cite{2018SR4}.

This is a core-collapse SNR with central PWN, so we use an ejecta mass of 5 M$_{\odot}$.
s=0 models for both fs emission and rs emission are all in the stage with $t_{core}<t<t_{rev}$.

The s=0, n=7, fs emission model gives age of 2200 yr and explosion energy of  0.63$\times10^{51}$ erg.
The s=0, n=9,  fs emission model gives age 2500 yr and explosion energy  0.48$\times10^{51}$ erg, but 
has the $EM$ of the rs larger than for the fs, so is not consistent.

The s=0, n=7, rs emission model has age of 1600 yr and explosion energy of 1.39$\times10^{51}$ erg,
but has $EM$ of the fs larger than for the rs, so is not consistent.
The s=0, n=9,  rs emission model has age of 880 yr and explosion energy of 3.43$\times10^{51}$ erg.
This model has $EM$ of the fs smaller than for the rs, so is consistent.
The s=0, n=12,  rs emission model has a low age (600 yr) and a large explosion energy (6.1$\times10^{51}$ erg) so is considered less likely.

s=2 models require large energies ($\sim$ 2$\times10^{52}$ erg) and low ages ($\sim$360 yr), so are considered unlikely.
Thus either s=0, n=7 fs emission or s=0, n=9 rs emission are consistent with the X-ray observations.
We have a preference for the fs emission model based on the consistency with the pulsar spin-down age, so we
adopt that one.

\section{Results and Discussion}\label{sec:properties}

\subsection{Comparison with Analytic ST Models}

We have applied our SNR model (extension of the TM models) to our SNR sample.
For comparision, we apply simpler ST models to find the corresponding  Sedov age,  ISM density and explosion energy  parameters. 
One procedure is to assume an explosion energy and ISM density to determine Sedov ages using the ST equations, e.g from \citet{1972Cox}.
Another approach is to use a model emission measure to derive density, then apply the ST 
equations to derive explosion energy and age from density and outer shock radius.
The latter method is used here. 

We compare two ST model calculations: 
one (commonly used) method uses a simple model emission measure and electron temperature equal to shock temperature (called simple model);
and the second uses emission measure and temperature profiles from the ST hydrodyamic solution combined with the Coulomb electron heating formalism 
(called electron heating model).
The simple model emission measure used is $dEM$=1\footnote{
%Most previous application of ST models to estimate SNR ages do not state their assumed emission measure or density distribution. 
One model density distribution, given by \citet{1998ApJ...505..732H} in their phenomenological model, %(their Section 3.1), 
is a uniform shell filling 1/4 of the SNR volume.
This model has a $dEM$ of $\pi/3$, similar to the one in our simple model.
}.
The electron heating model is similar to that used for the Sedov model of \citet{1998ApJ...505..732H} (their Section 3.2), 
although our model is simplified in the following way.
The electron heating model used here is self-similar, so we omit non-equilibrium ionization, which breaks self-similarity.
Electron heating is more important for SNR dynamics than non-equlibrium ionization (e.g. \cite{2013Ghav}), although non-equlibrium ionization is important for radiative losses,
and for deriving electron temperatures from observed X-ray spectra.
The electron temperatures used here (in Table~\ref{tab:TBLobserved}) are from  X-ray spectral models which include non-equilibrium ionization.
For the SNRs here, the ages are low enough that radiative losses are not important, so non-equlibrium ionization effects on the SNR models is expected to be small. 

We show our calculated results from both ST models in Table~\ref{tab:TBLST} using the input $EM$ and $kT$ and radius values from Table 1.
Consistent with the conclusion of \citet{1998ApJ...505..732H}, we find the observed electron temperature corresponds to a higher ion temperature in the
electron heating models. This implies a higher shock speed and thus a younger age for the electron heating models. 
Partially compensating for this is a second effect: 
the simple model takes the observed electron temperature to be the shock temperature; but the electron-heating
model uses the average over the entire shocked gas in the SNR, which is hotter than the shock temperature. 
This leads to a lower inferred shock temperature in the electron heating model for a given observed temperature. 
Thus older or higher density SNRs, where electron heating has time to equilibrate electron-ion temperatures, have a lower inferred shock speeds and increased ages. 
%Next we discuss the results from applying our full model, including effects of ejecta and resulting non-self-similar evolution phases.

The SNR radius and ISM density yield age and explosion energy from the ST formulae.
The ST solution assumes an ejected mass of zero and thus the emission is 100\% from the forward-shocked ISM. 
In cases where the assumptions of the ST model are approximately valid, i.e. for older SNRs, the ST
parameters are close (within $\sim10\%$) to those from the more complete models here. 
Often, some of the assumptions of the simple ST model are strongly violated. For example, the observed emission can
have a significant rs component, the ejecta mass can be large enough to strongly affect the evolution, or
the explosion can be in stellar wind environment instead of a uniform ISM. 
For those cases that the ST assumptions are not valid, we find  the Sedov age, ISM density and explosion energy  parameters
are often very different from those found using the more complete model that we have used here.
This demonstrates the need for using a more physically realistic model than the ST model.

\subsection{Comparison of TM with numerical ST model}

For validation, we compare our extended TM model results with results from another sophisticated model for the same set of SNRs.
A set of 7 SNRs in the Large Magellanic Cloud has been modelled using a numerically integrated Sedov model by \citet{1998ApJ...505..732H},
 and by \citet{2017Leahy} using the TM model. 
 \citet{1998ApJ...505..732H} used a Sedov temperature and emissivity profile, and integrated the non-equilibrium ionization equations for sets of
parameters to fit ASCA spectra of the SNRs. This was done for two cases (equal electron-ion  temperatures and Coulomb equilibration of electrons with ions).
\citet{2017Leahy} used the TM (s=0, n=7) shock evolution model, with a fixed ejecta mass of 1.4 M$_{\odot}$\footnote{
%In addition to fixing the parameters s=0, n=7 and M$_{ej}$=1.4M$_{\odot}$,
The \citet{2017Leahy} model is the same as the current TM model, but did not include calculation of the temperature and 
emission measure of rs material.}. 
The input data for the two models were different. \citet{1998ApJ...505..732H} used  ASCA data from 1993-1995, whereas \citet{2017Leahy} used
higher quality XMM-Newton results \citep{2016Maggi} from 2001-2015 and updated values of SNR radius. %derived from the best available images available in 2016.
 
The model values of $E_0$, $n_0$ and age from the two studies are compared in Table~\ref{tab:TBLlmcHL}.
It is seen that the input SNR radius is quite different for the two studies, especially for SNR N23. 
In order to compensate for this, our TM model was applied to these SNRs using the XMM-Newton $kT$ and $EM$ values 
but radius values from \citet{1998ApJ...505..732H}.
Those results are labelled as (L2) in Table~\ref{tab:TBLlmcHL}
are much closer to the  \citet{1998ApJ...505..732H} model values for most cases. This shows that the radius difference was the
main cause for the differences in derived parameters. 
Another cause of the parameter differences is the difference in SNR spectra observed by ASCA and XMM-Newton.
%\footnote{Data analysis of the ASCA observations, spectral fitting and fitting of the current model to the ASCA spectra is beyond the scope of the current paper.}.
Overall, the comparison shows that the \citet{1998ApJ...505..732H} model and current model produce similar results for the same SNRs.

\subsection{Comments on SNRs in Non-uniform Media}

Three of the SNRs above are mixed morphology (MM) SNRs: 3C391 (G31.9+0.0), W44 (G34.7-0.4), 
and W49B (G43.3-0.2). A model proposed for mixed morphology SNRs is the self-similar White and Long(WL) model for a SNR in a cloudy ISM \citep{1991WL}.
This model assumes zero mass of ejecta, but can be a good approximation to MM SNRs if
the age is large enough that the X-ray emission is dominated by shocked ISM and evaporating clouds compared to ejecta.
3C391 has EM dominated by shocked ISM. %and a smaller EM component from shocked ejecta with high Ca abundance; 
 W44  has a slight enhancement of  Si, S, Ar and Ca abundances so its emission is mainly ISM with a small contribution ($\sim$1\%) from ejecta.
W49B has higher iron abundance (5 times solar) but the emission is still mostly from the ISM (see estimate of swept-up mass in Section 3.1.13).
Thus the WL model should be a reasonably good approximation for these three SNRs.
Table~\ref{tab:TBLfullmodel} shows the results of applying the WL model to MM SNRs. The derived ages and explosion energies are the
same as for the TM model, supporting the TM model results. The WL model densities, which are intercloud densities, are lower
than the TM model densities,  which represent mean densities, by a factor of $\simeq$0.57 for all 3 MM SNRs.

These same MM SNRs have a detected recombining plasma (RP).
The origin of RP in SNRs is not yet well understood. 
There are two proposed explanations: the rarefaction model  in which shocked material breaks out from
dense circumstellar medium into rarefied interstellar medium \citep{2012Shimizu}, and the thermal conduction model 
in which evaporating dense material rapidly cools shocked plasma \citep{2011ZRP}. 
The latter is more consistent with the fact that MM SNRs are associated with detections of RP. 
This also indicates that the WL models are likely a better approximation than the TM models for MM SNRs.  
%Our results in Table~\ref{tab:TBLfullmodel} show that TM and WL models for these 3 SNRs give the same ages and explosion energies 
%but lower densities for the WL models compared to the TM models.

We applied both uniform ISM (s=0) and stellar wind (s=2) SNR models to each SNR. 
There was an expectation that a significant fraction of the SNRs would better be described by the s=2 models. 
CC SNR have different types of progenitors. \citet{2011Smith} (Fig. 1) give the fractions of SN types.
To obtain consistency with stellar evolution, \citet{2011Smith} argue that binary evolution is 
important and they give possible scenarios. % (their Figs. 5, 6, 7 and 8). 
We adopt their favored scenario (their Fig. 7), which has the following SN progenitors.
48.2\% are type II-P with initial mass 8.5-18.7 M$_{\odot}$. 6.4\% are type II-L with initial mass 18.7-23.1 M$_{\odot}$.
8.8\% are type II-n with initial mass  above 23.1 M$_{\odot}$. 27.7\% are type IIb + Ibc with initial mass above 8.5 M$_{\odot}$. 
8.8\% are type Ic with initial mass above 23.1 M$_{\odot}$.
The type II-P, II-L and II-n  (total of 63.4\% of SN) retain their hydrogen envelopes. %48.2+6.4+8.8=63.4
The other types are stripped-envelope SN via stellar winds, LBV eruptions, or Roche-lobe overflow in binary systems. 
A high fraction of the stellar wind systems and some fraction of the LBV and Roche-lobe systems are expected to have an
s=2 like circumstellar environment. This leads to the expectation that $\sim$20-40\% of CC SNR would be described by s=2 SNR models.
For type Ia SNR, single degenerate models are expected to have s=2, caused by the system mass-loss wind of the accreting white dwarf.
If the type Ia is caused by a double degenerate system, no wind is expected so s=0 is expected.

We find that only 2 of the 15 SNRs are best described by an s=2 model.
One of these is type Ia and the other is CC type.
The 6 unknown types in the sample (Table~\ref{tab:TBLobserved}) are all fit best with an s=0 model. 
Among those, there are likely one or two type Ia based on type Ia and CC SN frequencies  \citep{1991van}.
Those could be double degenerate type Ia SNR, thus consistent with s=0.
We find 7 of 8 confirmed CC type SNR are s=0 SNR.
Most massive stars, above $\sim$23 M$_{\odot}$  as noted above, are expected to have a stellar wind, resulting in s=2 SNR.
A significant fraction of massive stars are in binaries, where the interaction between the stars can complicate the circumstellar
density profile. Many massive stars are in stellar clusters where winds from nearby (fractions of a pc) massive stars
can interact with the wind of the progenitor and produce complex structures in the ISM which may be better represented
by a uniform density than a stellar wind profile.
Unless we are ready to accept the high explosion energies ($\sim 10^{53}$ erg) and low ages (few hundred yr) for the majority of s=2 models,
the current results are telling us that s=2 wind profiles are more rare than expected: $\sim$15\% compared to $\sim$20-40\% .

There are uncertainties introduced by applying spherically symmetric (SS) models to SNRs, because SNRs are observed to be non-symmetric. 
One measure of departure from SS is the departure of the major-axis to minor-axis ratio, $r_{axes}$ from 1.
For the SNRs in our sample, $r_{axes}$ 
has a range from 1.0 to 2.07, 
a mean of 1.32 and standard deviation of 0.32.
The most non-circular SNRs are G29.7-0.3, G32.8-0.1 and G41.1-0.3. % with $r_{axes}$ of 1.7 to 2.1.
G29.7-0.3 \citep{2009Su} and G32.8-0.1 \citep{2011Zhou} are interacting with molecular clouds and are X-ray and radio bright on one side.
G41.1-0.3 \citep{2005Safi} is elongated so has an asymmetric environment but has not been confirmed to be interacting with molecular cloud \citep{2018SR4}.
The departures from SS for non-symmetric SNRs will affect their evolution, and parameters determined by applying SS SNR models
will be approximations.

Existing SNR models are SS, so they cannot be used to determine the errors in using SS models for non-symmetric SNRs. 
\citet{2017LeahyWilliams} review analytic models for SNRs, which are all SS.
 \citet{1998ApJ...505..732H} and \citet{2001Bork} develop the numerical implementation of the ST model, which is SS, 
 and include non-equilibrium ionization for calculation of synthetic X-ray spectra.
\citet{2006Badenes}, and references therein, apply SS hydrodynamic SNR models with ejecta, combined with calculation of synthetic X-ray spectra to model X-ray
observations. 

Two possible ways of estimating errors in using SS models for non-symmetric SNRs are as follows. 
One method is to analyze a non-symmetric SNR in angular sectors, applying a different SS model to each sector. 
This assumes that the properties in a given sector do not affect those in neighboring sectors.
Then one can test  if the derived SNR properties differ by sector, and whether the summed properties from the sectors are consistent with
those derived from a single SS model applied to the whole SNR.
For example, one sector of an SNR expanding into a high density region or molecular cloud could be approximated by having a high density
model for that sector.
A second method is to create idealized asymmetric density distributions, and parametrize these in order to have a manageable set of models.
Then one can calculate the resulting SNR structure using a hydrodynamic code,  and calculate the resulting X-ray spectrum and image.
Finally one can fit the spectrum and image to data to obtain model parameters to compare to SS model parameters. 
These are goals for future work.

\subsection{Estimated SNR Population Properties}

The summary of our adopted models for the 15 SNRs is presented in  Table~\ref{tab:TBLfullmodel}. 
The statistical properties of the ages, explosion energies and ISM densities of the SNR sample are now examined.
A similar analysis was carried out for the LMC SNR population by \cite{2017Leahy}. For that
sample of 50 SNRs the observations are much closer to complete.
The current sample is limited to the part of the Galaxy covered by the VGPS survey and limited to SNRs with adequate X-ray spectra.
The volume of the Galaxy covered by the VGPS and by the X-ray observations is not easy to determine, partly because
of confusion limits and partly because of sensitivity limits, both in the presence of a variable background. 
For X-ray measurements, SNRs which are further from the Sun suffer more from absorption by Galactic
foreground material. 
One crude approximation of the fraction of the Galaxy observed in the current sample is the VGPS longitude range 
(18$^\circ$ to 67$^\circ$) divided by the full range of 360$^\circ$, or 0.136. 
This ignores missing SNRs which are at higher latitudes (the VGPS maximum latitude varies from 1.3 to 1.9$^\circ$),
but also ignores the concentration of SNRs towards smaller longitudes 
($\sim270^\circ$ to $360^\circ$ and $0^\circ$ to $\sim90^\circ$).
We compare to Green's catalog \citep{2017Green}, which lists 295 Galactic SNRs. 
49 of these are in the VGPS survey area, yielding a fraction of 0.166, similar to the estimate above.

The model ages give information on the SN birth rate, subject to the uncertainties of the incompleteness of the sample. 
Fig. 1 show the cumulative distribution of model ages for our sample of 15 SNRs. The straight lines are the expected
distributions for constant birth rates of 1 per 300 yr and 1 per 700 yr, respectively. The 1 per 300 yr line fits well the
SNRs with the smallest 3 ages and the 1 per 700 yr line fits approximately the SNRs with the 10 lowest ages.
This can be interpreted as increasing incompleteness of the sample for older SNRs. 
This is expected, as young SNRs are brighter and more easily measured to have an X-ray spectrum.
The expected SNR birth rate in the Galaxy is 1 per 40$\pm 10$ yr  \citep{1994Tammann}, with  rates per SN type (Ia, Ib and II) discussed by  \citet{1991van} (see Table 11 in
that paper).
Using the fraction above of the fraction of 0.166 of Galactic SNRs in the sample region this translates to 1/(200 yr) to 1/(300 yr) for the sample region.
This is close to the 1/(300 yr)  birth rate estimate shown in Fig. 1 for the youngest SNRs (age $<$1000 yr). 
The agreement suggests that the X-ray detections are near 100\% for the youngest SNRs.
It also suggests that the decreasing birth rate with increasing age seen in Fig. 1 is caused by increasing 
incompleteness of X-ray (and perhaps radio) detections for older SNRs. 
I.e. for ages $\lesssim$5000 yr the incompleteness is a factor $\sim$2, yielding half the expected birth rate.
The incompleteness continues to increase for higher ages, reaching a factor $\sim$4 for the 
oldest SNRs ($\sim$15000 yr) in the sample .

The cumulative distribution of explosion energies is shown in Fig. 2. This distribution is not consistent with 
uniform or Gaussian distributions, but is consistent with a log-normal distribution. 
The property that SNR explosion energies follows a log-normal distribution was discovered by \citet{2017Leahy}.
The solid line in Fig. 2 is the best fit cumulative distribution, with parameters  
$E_{av}= 5.4\times10^{50}$ erg and $\sigma_{logE}$=0.45, which corresponds
to a 1-$\sigma$ dispersion factor of 1.9 in energy.
These parameters are similar to as those derived from the LMC SNR sample of 
$E_{av}= 5.4\times10^{50}$erg and $\sigma_{logE}$=0.47  \citep{2017Leahy}.
This suggests that the LMC and Galactic SNR samples are very similar, and thus LMC and Galactic SN are very similar.

The cumulative distribution of ISM densities is shown in Fig. 3 for the 13 of the 15 SNRs with a derived ISM density. 
This distribution is consistent with a log-normal distribution.
The property that ISM densities in the neighborhood of SN explosions has  a log-normal distribution was 
first shown by \citet{2017Leahy}.
The best fit cumulative distribution is shown by the solid line in Fig. 3. 
The parameters  are $n_{0,av}= 0.26$ cm$^{-3}$ and $\sigma_{log(n_0)}$=0.80, which corresponds
to a 1-$\sigma$ dispersion factor of 6.3 in density\footnote{Here we used our TM model (s=0) densities  for the three MM-type SNRs. 
If we use the WL model densities for the three MM-type SNRs, we obtain $n_{0,av}= 0.23$ cm$^{-3}$ and and $\sigma_{log(n_0)}$=0.74.}.
For the LMC SNR sample \citep{2017Leahy}, the mean density was considerably lower, $n_{0,av}= 0.08$ cm$^{-3}$
and the dispersion similar.
The difference in mean density is not surprising because the current sample is mainly SNRs in the inner Galaxy whereas the LMC sample
was for the whole LMC. 

\section{Conclusion}\label{sec:conclusion}

Distances to Galactic SNRs have improved significantly, allowing determination of radii and enabling 
the application of SNR models.
We extended spherically symmetric SNR evolution models to include effects of ejecta mass and emission from shocked ejecta.
We applied the models to estimate SNR parameters for our sample of SNRs from the inner Galaxy. 
Then we examined the distributions of the parameters to determine properties of the Galactic SNR population in the inner Galaxy.
The estimated birth rate is consistent with estimates of the overall Galactic SNR birth rate
and with estimates of the fraction of SNRs in the Galaxy in our sample.
We find that the energies and ISM densities of SNR can be well fit with log-normal distributions. 
The distribution of explosion energies is very similar to that for SNRs in the Large
Magellanic Cloud (LMC), suggesting a surprisingly close similarity in the population of SN explosions in the LMC and in the Galaxy.
The ISM density distributions for Galactic and LMC SNRs have similar dispersion but Galactic SNRs have a higher mean density by
factor $\sim2.5$. 
A higher mean density is expected because our sample SNRs are selected from the inner Galaxy.

In future, we plan to extend this type of study of SNRs to include all Galactic SNRs with measured distances. 
The goal is to significantly increase the sample size and better determine the intrinsic properties of SNRs.

\acknowledgments
This work was supported by a grant from the Natural Sciences and Engineering Research Council of Canada. 
The authors thank the referee and editor for providing useful comments which improved this work.

\clearpage

%% the file, uncomment the \plotone, \plottwo, and
%% If you need a layout that cannot be achieved with \plotone or
%% \plottwo, you can invoke the graphicx package directly with the
%% \includegraphics command or use \plotfiddle. For more information,
%% please see the tutorial on "Using Electronic Art with AASTeX" in the
%% documentation section at the AASTeX Web site, http://aastex.aas.org/
%%
%% The examples below also include sample markup for submission of
%% supplemental electronic materials. As always, be sure to check
%% the instructions to authors for the journal you are submitting to
%% for specific submissions guidelines as they vary from

%% This example uses \plotone to include an EPS file scaled to
%% 80% of its natural size with \epsscale. Its caption
%% has been written to indicate that additional figure parts will be
%% available in the electronic journal.

%\begin{figure}
%\epsscale{.50}
%\plotone{NewImage1.eps}
%\caption{caption.\label{fig1}}
%\end{figure}
%\clearpage

\begin{figure*}[ht!]
\includegraphics[width=\textwidth]{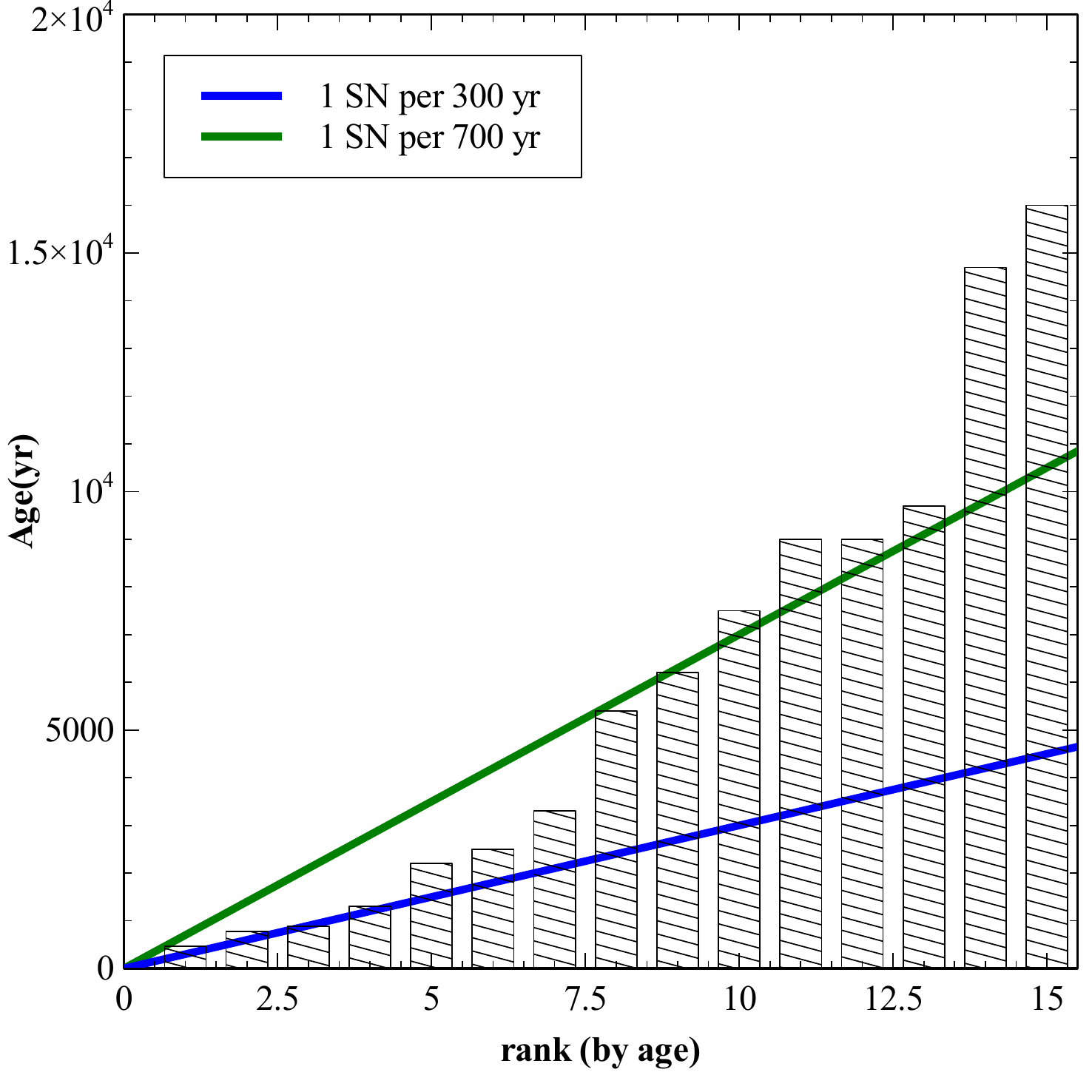}
\caption{Cumulative distribution of model ages and fit lines for birth rate of 1 per 300 yr and 1 per 700 yr.}
\label{fig:1}
\end{figure*} 

\begin{figure*}[ht!]
\includegraphics[width=\textwidth]{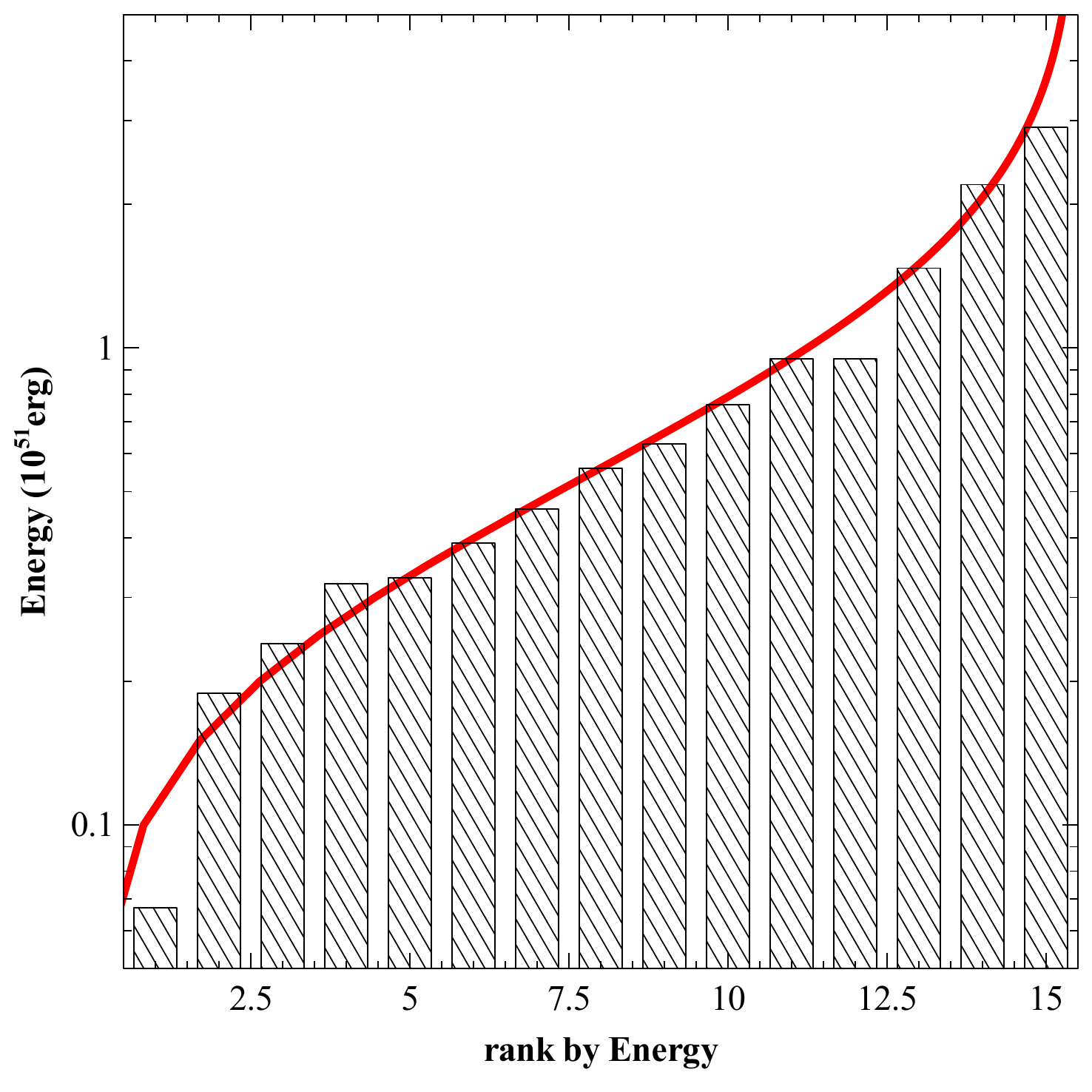}
\caption{Cumulative distribution of model  explosion energies of the 15 SNRs and fit line for a log-normal distribution of explosion energies with mean
explosion energy of E$_0$=5.4$\times10^{50}$ erg and variance in log(E$_0$) of 0.45.}
\label{fig:2}
\end{figure*} 

\begin{figure*}[ht!]
\includegraphics[width=\textwidth]{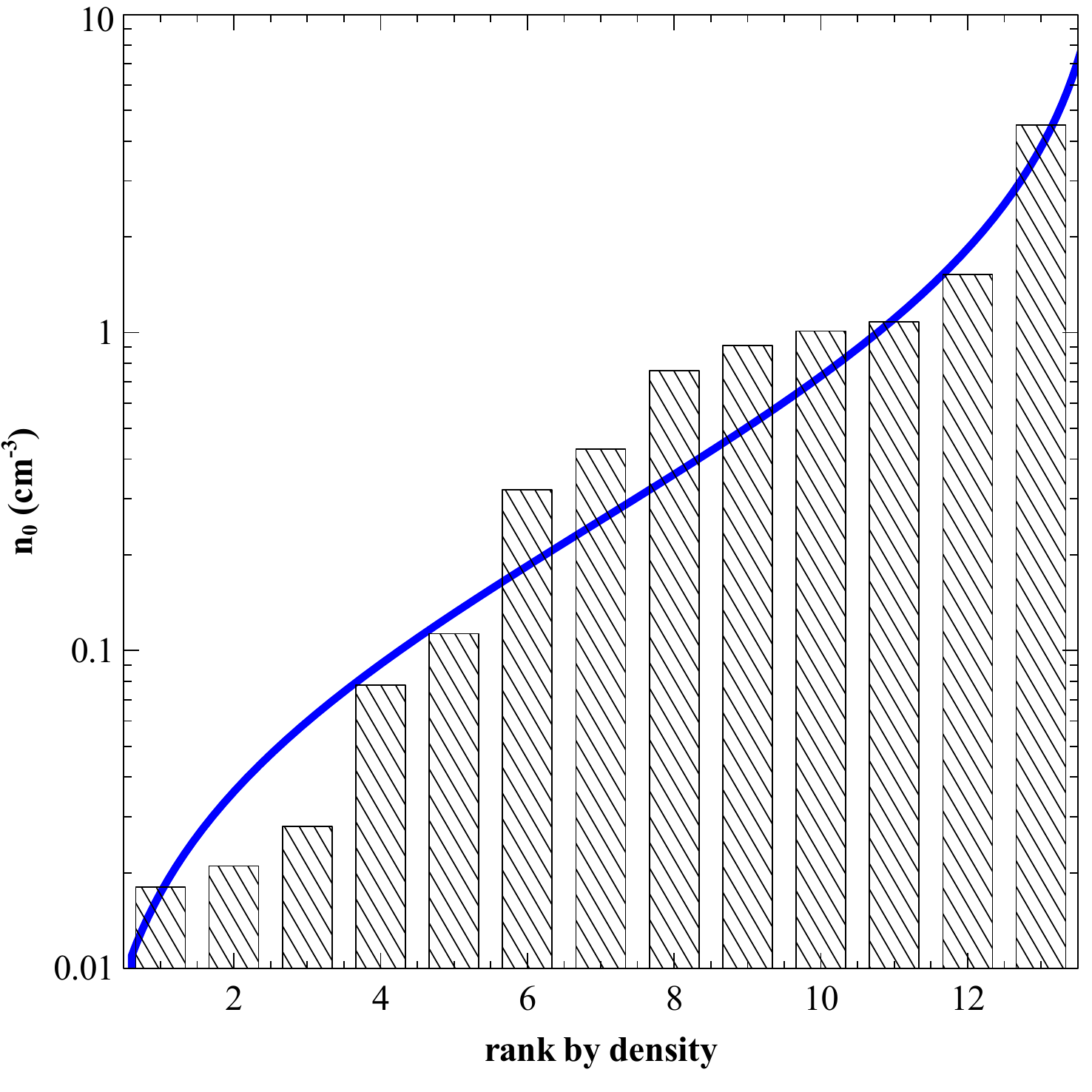}
\caption{Cumulative distribution of model ISM densities and fit line for a log-normal distribution of ISM densities with mean
density of n$_0$=0.26 cm$^{-3}$ and variance in log(n$_0$) of 0.80.}
\label{fig:3}
\end{figure*} 

\clearpage

\clearpage

\begin{deluxetable}{crrrrrrrrrr}
\tabletypesize{\scriptsize}
%\rotate
\tablecaption{SNR Observed Quantities} %$^{a}$
\tablewidth{0pt}
\tablehead{
\colhead{SNR} & \colhead{Type}  & \colhead{Distance} & \colhead{Semi-axes}  &  \colhead{Radius}      & \colhead {$EM^{a,b}$}     & \colhead {$kT^{a,b}$} & \colhead {$EM_2^{a,b}$}     & \colhead {$kT_2^{a,b}$} \\
\colhead{}      & \colhead{}           & \colhead{(kpc)}   & \colhead{(arcmin)}   & \colhead{(pc)}      & \colhead {($10^{58}$cm$^{-3}$)} & \colhead {(keV)} & \colhead {($10^{58}$cm$^{-3}$)} & \colhead {(keV)}\\
% & \colhead{(yr)} 
}
\startdata
 G18.1-0.1  &  ?   &   6.4$\pm$0.2 & 4.0$\times$3.6 & 7.0$\pm$0.2  &   8.3(2.9-21)d$_{6.4}^2$  & 0.44(0.37-0.56) & n/a & n/a \\   
 G21.5-0.9  &  CC   &   4.4$\pm$0.2 & 1.8$\times$1.5 &  2.1$\pm$0.1  &   1.20(0.51-3.4)d$_{4.4}^2$ & 0.33 (0.25-0.45) & n/a & n/a \\   
 G21.8-0.6  &  ?   &   5.6$\pm$0.2  & 10$\times$10 &  16.3$\pm$0.6   & 0.75(0.077-0.83)d$_{5.6}^2$ & 0.62(0.60-0.64)& n/a & n/a  \\  
G27.4+0.0  &  CC   &   5.8$\pm$0.3 & 2.9$\times$2.6 &  4.6$\pm$0.25    & 5.4(1.44-14.3)d$_{5.8}^2$ &1.60(0.90-2.4)& 49(37-68)d$_{5.8}^2$ & 0.50(0.30-0.60) \\
G28.6-0.1  &  ?   &   9.6$\pm$0.3  & 12$\times$9.0&  29.3$\pm$0.9   &  0.58(0.30-1.2)d$_{9.6}^2$ & 0.71(0.55-0.93) & n/a & n/a \\ 
G29.7-0.3  &  CC  &   5.6$\pm$0.3 & 2.6$\times$1.5 &  3.3$\pm$0.15   & 0.86(0.43-1.29)d$_{5.6}^2$ & 0.70(0.60-0.80)  & n/a & n/a \\ 
G31.9+0.0  & CC  &   7.1$\pm$0.4  & 4.0$\times$3.0 &  7.2$\pm$0.4   &  220(150-330)d$_{7.1}^2$ & 0.17(0.16-0.18) & 2.1(0.4-2.3)d$_{7.1}^2$ & 0.49(0.48-0.51) \\
G32.8-0.1  & ?  &   4.8$\pm$0.3 & 12.4$\times$6.0 &  13.0$\pm$0.8  & 0.028(.014-.069)d$_{4.8}^2$ &  0.65(0.44-0.97) & n/a & n/a \\ 
G33.6+0.1  & CC  &   3.5$\pm$0.3& 6.3$\times$5.9  &  6.3$\pm$0.4  &  2.9(2.5-4.5)d$_{3.5}^2$ & 0.22(0.20-0.25) & 0.50(.43-.57)d$_{3.5}^2$ & 0.84(0.79-0.89) \\ 
G34.7-0.4  & ?  &   3.0$\pm$0.3 & 17.7$\times$12 &  13.0$\pm$1.3   & 146(132-162)d$_{3.0}^2$ & 0.49(0.43-0.54)  & n/a   &  n/a \\ 
G39.2-0.3  & CC  &   8.5$\pm$0.5 & 3.9$\times$3.5 &  9.1$\pm$0.5  &  2.8(0.91-5.0)d$_{8.5}^2$ & 0.68(0.49-1.05) & 0.94(.74-1.14)d$_{8.5}^2$ & 1.21(1.08-1.34) \\ 
G41.1-0.3  & Ia  &   8.5$\pm$0.5  & 2.3$\times$1.3 &  4.45$\pm$0.25  & 950(650-1380)d$_{8.5}^2$  & 0.21(0.20-0.23) & 0.99(.81-1.18)d$_{8.5}^2$ &  1.50(1.30-2.0) \\ 
G43.3-0.2  & ?  &   11.3$\pm$0.4 & 2.6$\times$2.6 &  8.55$\pm$0.3   &  24.6(24.4-24.8)d$_{11.3}^2$ & 1.52(1.50-1.53) & n/a & n/a  \\ 
G49.2-0.7  & CC  &  5.6$\pm$0.6 & 18.5$\times$18.5 &  29$\pm$3.3   &  0.82(0.71-1.01)d$_{5.6}^2$ &  0.70(0.65-0.76) & n/a & n/a \\ 
G54.1+0.3  & CC  &  4.9$\pm$0.8& 6.6$\times$4.5  &  7.9$\pm$1.3   & 0.26(0.125-0.39)d$_{4.9}^2$ & 2.0(1.60-2.4) & n/a  & n/a \\ 
\enddata
\label{tab:TBLobserved}
\tablenotetext{a}{For SNRs with only one measured thermal plasma component, $EM$ and $kT$ are given; for SNRs with two measured thermal plasma components,
$EM$ and $kT$ are given for the first component and $EM_2$ and $kT_2$ are given for the second component.}
\tablenotetext{b}{For X-ray spectra of SNRs, the errors in $EM$ and $kT$ are usually not symmetric, thus the quantities in parentheses are the 1$\sigma$ lower
and upper limits.}
\end{deluxetable}
\clearpage

\begin{deluxetable}{crrrrrrrrrrrr}
\tabletypesize{\scriptsize}
%\rotate
\tablecaption{Basic and Electron-heating ST Model Results} %$^{a}$
\tablewidth{0pt}
\tablehead{
& \multicolumn{3}{c}{Basic ST model} & \multicolumn{3}{c}{ST with electron heating} \\
%\colhead{} & \colhead{Simple} &  \colhead{ }    & \colhead{ }  & \colhead{Heating} &  \colhead{ }    & \colhead{ }  \\
\colhead{SNR} & \colhead{Age} &  \colhead{Energy}    & \colhead{$n_0$}  &  \colhead{Age} &  \colhead{Energy}   & \colhead{$n_0$} \\
\colhead{}    & \colhead{(yr)}  & \colhead{($10^{51}$erg)}  & \colhead{(cm$^{-3}$)} & \colhead{(yr)}  & \colhead{($10^{51}$erg)}  & \colhead{(cm$^{-3}$)} \\
}
\startdata
 G18.1-0.1  &   9000 & 0.175 & 0.66 &  5100 & 0.189 & 0.92 \\   
 G21.5-0.9  &   1550 & 0.0082 & 1.53 &   1800 & 0.0089 & 2.1 \\   
 G21.8-0.6  &    8800 & 0.26 & 0.056 &   9200 & 0.34 & 0.078 \\  
G27.4+0.0  &     1560 & 0.28 & 0.99 &   1580 & 0.37 & 1.37 \\ 
G28.6-0.1  &    14800 & 0.64 & 0.020 &   14200 & 0.96 & 0.028 \\ 
G29.7-0.3  &   1690 & 0.030 & 0.65 &   1850 & 0.034 & 0.90  \\ 
G31.9+0.0  &   7400 & 0.36 & 3.3 &   8400 & 0.39 & 4.5 \\ 
G32.8-0.1  &  7000 & 0.037 & 0.015 &   5700 & 0.077 & 0.021 \\ 
G33.6+0.1  &   5700 & 0.044 & 0.46  &   6500 & 0.048 & 0.64 \\ 
G34.7-0.4  &  7900 & 2.1 & 1.09 &   8900 & 2.2 & 1.54 \\ 
G39.2-0.3  &    4700 & 0.23 & 0.26 &   5200 & 0.27 & 0.36 \\ 
G41.1-0.3  &   4100 & 0.45 & 13.9  &   4700 & 0.49 & 19.4 \\ 
G43.3-0.2  &   2900 & 1.41 & 0.84  &   3200 & 1.70 &1.17 \\ 
G49.2-0.7  &    16700  & 0.89 & 0.020 &  16400 & 1.29 & 0.028 \\ 
G54.1+0.3  &    2400  & 0.17 & 0.010 &  1500 & 0.56 & 0.136 \\ 
\enddata
\label{tab:TBLST}
%\tablenotetext{a}{.}
\end{deluxetable}

\clearpage
\begin{deluxetable}{crrrrrrrrrrrr}
\tabletypesize{\scriptsize}
%\rotate
\tablecaption{Comparison of TM Model$^{a}$ and NEI Sedov model$^{b}$ for LMC SNRs} %$^{a}$
\tablewidth{0pt}
\tablehead{
%& \multicolumn{3}{c}{Basic ST model} & \multicolumn{3}{c}{ST with electron heating} \\
%\colhead{} & \colhead{Simple} &  \colhead{ }    & \colhead{ }  & \colhead{Heating} &  \colhead{ }    & \colhead{ }  \\
\colhead{SNR} &  \colhead{Radius$^{c}$}    &  \colhead{Energy$^{c}$}    & \colhead{$n_0$$^{c}$}  &  \colhead{Age$^{c}$}  \\
\colhead{}    & \colhead{(pc)}   & \colhead{($10^{51}$erg)} & \colhead{(cm$^{-3}$)} & \colhead{($10^{3}$ yr)}  \\
}
\startdata
J0505-6802 (N23) & 6.7(H) 11.6(L) & 0.46,0.53(H) 1.06(L),0.46(L2) &  1.6,1.7(H) 1.5(L),3.5(L2) & 3.8,3.6(H) 10.0(L),5.8(L2)  \\
J0526-6605 (N49) &  8.2(H) 10.2(L) & 1.5,1.6(H) 1.2(L),0.84(L2) &  2.6,2.6(H) 1.9(L),2.7(L2)  & 4.4,4.3(H) 7.6(L),6.2(L2)  \\
J0535-6602 (N63A) &  8.5(H) 9.80(L) &  2.6,2.9(H) 2.0(L),1.6(L2) &  3.9,3.9(H) 3.0(L),3.8(L2)  &  4.5,4.2(H) 6.6(L),5.7(L2) \\ 
J0505-6753 (DEM 71) &  10.4(H) 9.4(L) & 1.1,1.4(H) 0.67(L),0.77(L2) &  0.67,0.86(H) 1.28(L),1.09(L2) & 4.7,4.7(H) 6.8(L),7.5(L2) \\
J0525-6938 (N132D) &  12.1(H) 13.4(L) & 6.1,5.7(H) 4.8(L),4.1(L2) &  2.5,2.7(H) 2.3(L),2.6(L2) & 5.4,6.7(H) 8.1(L),7.3(L2) \\
J0453-6829 (0453-68.5) &  15.0(H) 14.2(L) &  0.91,1.2(H) 0.83(L),0.90(L2) &  0.30,0.28(H) 0.58(L),0.53(L2) & 8.7,7.4(H) 11.4(L),12.0(L2)   \\
J0525-6559 (N49B) & 17.0(H) 18.9(L) & 2.7,3.1(H), 2.8(L),2.4(L2) &  0.75,0.82(H) 0.51(L),0.60 (L2) & 10.9,10.6(H) 11.8(L),10.7(L2)  \\
\enddata
\label{tab:TBLlmcHL}
\tablenotetext{a}{The TM model is from \citet{2017Leahy}. The model is nearly the same as the current SNR model: see text for explanation of differences.
 The $EM$ and $kT$ were from 2001-2015 XMM-Newton observations.}
\tablenotetext{b}{The NEI Sedov model is from \citet{1998ApJ...505..732H}. The data were from 1993-1995 ASCA observations.}
\tablenotetext{c}{(H) indicates values from \citet{1998ApJ...505..732H} with values given electron-ion equilibration and for Coulomb equilibration, respectively; 
(L) indicates values from \citet{2017Leahy}, (L2) indicates the TM model run with XMM-Newton data but with radii from \citet{1998ApJ...505..732H}.}
\end{deluxetable}

\clearpage

\begin{deluxetable}{crrrrrrrrrr}
\tabletypesize{\scriptsize}
%\rotate
\tablecaption{SNR Model Results} %$^{a}$
\tablewidth{0pt}
\tablehead{
\colhead{SNR} & \colhead {model type$^{a}$} & \colhead {s} & \colhead {n}  & \colhead {$M_{ej}$} & \colhead{Age} &  \colhead{Energy}    & \colhead{$n_0$(s=0)}    & \colhead{$\rho_s$(s=2)} \\
\colhead{}   & \colhead{(fs or rs)}  & \colhead{}      & \colhead{}   & \colhead{($M_{\odot}$)} & \colhead{(yr)}  & \colhead{($10^{51}$erg)}  & \colhead{(cm$^{-3}$)} & \colhead{(M$_{\odot}$s/(km~yr))} \\
% & \colhead {Sedov Age}    % & \colhead{(yr)} 
}
\startdata
 G18.1-0.1  &  fs &  0 &  7 & 1.4 & 5400 & 0.189 & 0.92 & n/a \\   
 G21.5-0.9  &   rs &  0 & 9 & 5 & 470 & 0.56 & 1.01 & n/a \\   
 G21.8-0.6  &   fs &  0 & 7 & 1.4 & 9700 & 0.33 & 0.078 & n/a \\  
G27.4+0.0  &   fs &  0& 12 & 10 &  2500 & 0.32 & 1.08 & n/a \\ 
G28.6-0.1  &    fs &  0  & 7 & 1.4  & 14700 & 0.95 & 0.028 & n/a \\ 
G29.7-0.3  &   rs &  0 &  9 & 5 & 890 & 0.46 & 0.43 & n/a \\ 
G31.9+0.0  &  rs &  0 &  9 & 10  & 9000 & 0.39 & 4.5 & n/a \\ 
                  & WL & n/a & n/a & n/a & 8700 & 0.39 & 2.6 & n/a \\
G32.8-0.1  &  fs &  0 &  7 & 1.4  & 7500 & 0.067 & 0.018 & n/a \\ 
G33.6+0.1  &  fs & 2 & 7 & 20  & 780 & 2.9 & n/a & 6.3$\times10^{-8}$ \\ 
G34.7-0.4  &  fs &  0 & 7 & 5  & 9100 & 2.2 & 1.52 & n/a \\ 
                  & WL & n/a & n/a & n/a & 9200 & 2.2 & 0.86 & n/a \\
G39.2-0.3  &   fs &  0 & 7 & 5 & 6200 & 0.24 & 0.32 & n/a \\ 
%G41.1-0.3  &  fs &  0 & 7 & 1.4  & 4700 & 0.49 & 19.4 & n/a \\ 
G41.1-0.3  &  fs &  2 & 7 & 1.4  & 1300 & 0.95 & n/a & 9.5$\times10^{-7}$ \\ 
G43.3-0.2  &  fs &  0 & 7 & 1.4 & 3250 & 1.69 & 1.17 & n/a \\ 
                  & WL & n/a & n/a & n/a & 3100 & 1.82 & 0.66 & n/a \\
G49.2-0.7  &   fs &  0 & 9  & 10 & 16000  & 0.76 & 0.021 & n/a \\ 
G54.1+0.3  &   fs &  0 & 7  & 5 &  2200  & 0.63 & 0.113 & n/a \\ 
\enddata
\label{tab:TBLfullmodel}
\tablenotetext{a}{Model type: fs is forward shock model; rs is reverse shock model ; WL is the model of \citet{1991WL} with Coulomb equilibration added.}
\end{deluxetable}
\clearpage

%% The following command ends your manuscript. LaTeX will ignore any text

\end{document}